\documentclass[a4paper,12pt]{article}

\pdfoutput=1

\usepackage{amssymb,amsmath,bm}
\usepackage{a4wide}
\usepackage{color}
\usepackage{slashed}
\usepackage{cite}
\usepackage{graphicx}
\usepackage[latin1]{inputenc}
\usepackage{amsfonts}
\usepackage{lscape}
\usepackage{amsthm}
\usepackage{booktabs}
\usepackage{array}
\usepackage{rotating}
\usepackage{float}
\usepackage{multirow}
\usepackage{graphicx,rotating,amsmath,multirow,xcolor,colortbl,xcolor}
\usepackage{hyperref,pdflscape,slashed}
\definecolor{rosso}{cmyk}{0,1,1,0.4}
\definecolor{rossos}{cmyk}{0,1,1,0.55}
\definecolor{rossoc}{cmyk}{0,1,1,0.2}
\definecolor{blu}{cmyk}{1,1,0,0.3}
\definecolor{blus}{cmyk}{1,1,0,0.6}
\definecolor{bluc}{cmyk}{1,1,0,0.1}
\definecolor{verde}{cmyk}{0.92,0,0.59,0.25}
\definecolor{verdec}{cmyk}{0.92,0,0.59,0.15}
\definecolor{verdes}{cmyk}{0.92,0,0.59,0.4}

\hypersetup{colorlinks,bookmarksopen,bookmarksnumbered,
linkcolor=blus,pdfstartview=FitH,urlcolor=rossos,citecolor=verde}

\font\tenrsfs=rsfs10 at 12pt
\font\sevenrsfs=rsfs7
\font\fiversfs=rsfs5
\newfam\rsfsfam
\textfont\rsfsfam=\tenrsfs
\scriptfont\rsfsfam=\sevenrsfs
\scriptscriptfont\rsfsfam=\fiversfs
\def\mathscr#1{{\fam\rsfsfam\relax#1}}

\definecolor{Gray}{gray}{0.95}

\usepackage[small,bf]{caption}
\renewcommand\Im{\operatorname{Im}}

\newcommand{\IM}{\Im \left(M_{12}\right) \,}

\newcommand{ \mysmall}[1]{\scriptscriptstyle #1} % a smaller #

\newcommand{\be}{\begin{equation}}
\newcommand{\ee}{\end{equation}}
\newcommand{\bea}{\begin{eqnarray}}
\newcommand{\eea}{\end{eqnarray}}
\newcommand{\beq}{\begin{equation}}
\newcommand{\eeq}{\end{equation}}
\newcommand{\beqa}{\begin{eqnarray}}
\newcommand{\eeqa}{\end{eqnarray}}
\newcommand{\nn}{\nonumber}

\newcommand{\eq}{\begin{equation}}
\newcommand{\eqx}{\end{equation}}
\newcommand{\eqn}{\begin{eqnarray}}
\newcommand{\bi}{\begin{itemize}}
\newcommand{\eqnx}{\end{eqnarray}}
\newcommand{\ei}{\end{itemize}}

\def\nn{\nonumber}

\newcommand{\g}{{\tilde g}}

\newcounter{hran}

\def\eq#1{eq.~(\ref{#1})}
\selectfont

\begin{document}

%\centerline{CERN-PH-TH-2015-191\hfill IFUP-TH/2015}

\vspace{0.2truecm}

\begin{center}
\boldmath

{\textbf{\Large\color{black} CP Violation Tests of Alignment Models at LHCII }} 
\unboldmath

\bigskip

%\vspace{0.4truecm}
\vspace{0.5truecm}

{\bf Diptimoy Ghosh$^a$, Paride Paradisi$^b$, Gilad Perez$^{a}$, Gabriele Spada$^c$}  \\[5mm]

{\it $^a$ Department of Particle Physics and Astrophysics, Weizmann Institute of Science, Rehovot 76100, Israel}\\[0mm]
{\it $^b$ Dipartimento di Fisica dell'Universit{\`a} di Padova and INFN, Italy}\\[0mm]
{\it $^c$ Scuola Internazionale Superiore di Studi Avanzati (SISSA) and INFN, Italy}\\[0mm]
\end{center}

%\vspace{-1cm}

\begin{quote}
%\large
\noindent

\centerline{\bf Abstract}
\vskip .1in
We analyse the low-energy phenomenology of alignment models both model-independently and within supersymmetric (SUSY) scenarios
focusing on their CP violation tests at LHCII. 
%Assuming that New Physics (NP) contributions to $K^0-\bar K^0$ and $D^0-\bar D^0$ mixings are approximately $SU(2)_L$ invariant, 
Assuming that New Physics (NP) contributes to $K^0-\bar K^0$ and $D^0-\bar D^0$ mixings only through non-renormalizable operators 
involving $SU(2)_L$ quark-doublets, we derive model-independent correlations among CP violating observables of the two systems.
Due to universality of CP violation in $\Delta F=1$ processes the bound on CP violation in Kaon mixing generically leads to an upper 
bound on the size of CP violation in $D$ mixing. 
Interestingly, this bound is similar in magnitude to the current sensitivity reached by the LHCb experiment which is starting now to probe 
the natural predictions of alignment models.
Within SUSY, we perform an exact analytical computation of the full set of contributions for the $D^0-\bar D^0$ mixing amplitude. 
We point out that chargino effects are comparable and often dominant with respect to gluino contributions making their inclusion 
in phenomenological analyses essential. As a byproduct, we clarify the limit of applicability of the commonly used mass insertion 
approximation in scenarios with quasi-degenerate and split squarks.
\end{quote}

\thispagestyle{empty}
\vfill

\vspace{-0.5cm}

%----------------------------------------------------------------------------
%\tableofcontents

%%%%%%%%%%%%%%%%%%%%%%%%%%%%%%%%%%%%%%%%%%%%%%%%%%%%%%%%%%%%%%%%%%
\section{Introduction}
%%%%%%%%%%%%%%%%%%%%%%%%%%%%%%%%%%%%%%%%%%%%%%%%%%%%%%%%%%%%%%%%%%

The meson systems are among the most interesting low-energy probes of New Physics (NP)
and can be regarded as golden channels of the high intensity frontier.
However, all the currently available data on $K$ and $B_{d,s}$ systems agree 
well with the Standard Model (SM) predictions. In turn, this leads to the so-called
NP flavor and CP puzzles, that is the tension between the solution of the hierarchy
problem, requiring a TeV scale NP, and the explanation of the flavor physics 
data.

One option to reconcile the above tension without giving up on naturalness is to assume that NP is flavor blind.
This could either arise when the flavor mediation scale is very high leading to minimal flavor violation~\cite{MFV,Kagan:2009bn}, 
or possibly when non-abelian flavor symmetries are involved~\cite{Dine:1993np}. In both cases, however, flavor non-universality 
effects involving the first two generations are suppressed, both in the luminosity and energy frontiers~(see, e.g.,~\cite{Kagan:2009bn,Grossman:2007bd,Gedalia:2010rj}).

However, another possibility regarding the flavor structure of NP might be realised in Nature. 
This is due to the fact that most of the information that we have involving low-energy, flavor violating, probes of the SM involve down 
type fermions. Thus, there is always the possibility that new physics is in fact at the TeV scale and yet it is aligned with the down type 
Yukawa matrices~\cite{Nir:1993mx,Leurer:1993gy,Nir:2002ah,Galon:2013jba,Delaunay:2013iia}. In such a case flavor universality in 
the first two generations is badly broken, leading to several interesting signatures at the LHC~\cite{DaRold:2012sz,Delaunay:2013iia,Delaunay:2013pwa,Mahbubani:2012qq}. 
Somewhat surprisingly such a framework might even be linked to the hierarchy problem leading to flavorful naturalness~\cite{Blanke:2013zxo}.

In all above cases, $D$ physics observables represent a unique tool to probe NP flavor
effects, quite complementary to tests in $K$ and $B$ systems. On general grounds, 
$D$ systems offer a splendid opportunity to discover CP violating effects arising from 
NP~\cite{Bergmann:2000id,Gedalia:2009kh,Grossman:2009mn,Bigi:2009df,Kagan:2009gb} 
as the SM predictions are expected to be of order ${\cal O}(V_{cb}^*V_{ub}/V_{cs}^*V_{us})\sim10^{-3}$.
As a consequence, any experimental signal of CP violation in $D^0-\bar D^0$ above the
per mill level would probably point towards a NP effect.

In this work, we revisit the phenomenology of alignment models model-independently as well as within SUSY scenarios.
Assuming that NP contributes to $K^0-\bar K^0$ and $D^0-\bar D^0$ mixings only through non-renormalizable operators 
involving $SU(2)_L$ quark-doublets,
%Assuming that NP contributions to $K^0-\bar K^0$ and $D^0-\bar D^0$ mixings are approximately $SU(2)_L$ invariant, 
we derive model-independent correlations among CP violating observables of the two systems. At this era of the beginning 
of the second run of the LHCb we can safely assume that CP violation effects in the $D$ system are small and thus many 
of the theoretical expressions are simplified, as we are allowed to work at the linear order in the CP violating parameters. 

We briefly summarise here our findings related to the model-independent analysis: 
\begin{itemize}
\item[{\bf i)}]
generically the bound on the allowed amount of CP violation in the Kaon system limits the possible size of 
CP violation in mixing in the $D$ system;
\item[{\bf ii)}] this bound is similar in magnitude to the current sensitivity reached by the LHCb experiment. As such,
a discovery of CP violation in $D$-mixing would be quite challenging for alignment (and many other) models;
\item[{\bf iii)}] the expected resolutions at the next LHCb run, as well as other potential experiments, will provide 
useful information on the parameter space of models where CP violation is controlled dominantly by the left-handed sector. 
\end{itemize}
%
%and then describe the result of our study of the corresponding observables within SUSY:  
Then, we focus on SUSY alignment models and the main goals of our study are:
\begin{itemize}
\item[{\bf i)}] to perform an analytical computation of all SUSY contributions
(pure gluino, mixed neutralino/gluino, chargino, as well as neutralino contributions)
for the $D^0-\bar D^0$ mixing amplitude;
\item[{\bf ii)}] to study the allowed ranges for the squark masses which are compatible
with both collider and flavor physics constraints;
\item[{\bf iii)}] to study the allowed effects for charm-CPV pointing out possible
correlations among $D$ and $K$ meson observables enabling to probe or falsify
the NP scenario in question;
\item[{\bf iv)}] to clarify the limit of applicability of the commonly used Mass 
Insertion (MI) approximation comparing the full and approximated results in two relevant 
squark mass regimes: the quasi-degeneracy and split scenarios.
\end{itemize}

Our paper is organized as follow: in sec.~\ref{sec:DDbar_observables} we review the main
formalism and formulae for $D^0-\bar D^0$ mixing observables.
In sec.~\ref{sec:mod_ind_analysis}, we derive model-independent correlations among CP 
violating observables related to $D^0$ and $K^0$ systems.
Sec.~\ref{sec:DDbar_align} is devoted to the calculation of the $D^0-\bar D^0$ mixing amplitude
in SUSY, while the study of charm-CP violation is presented in sec.~\ref{sec:CPV_mixing}. 
Our main results and conclusions are summarized in sec.~\ref{sec:conclusions}. 
Finally, in Appendix A and B we specify the notation used in the text and report the loop functions,
respectively.

%%%%%%%%%%%%%%%%%%%%%%%%%%%%%%%%%%%%%%%%%%%%%%%%%%%%%%%%%%%%%%%%%%
\section{$D^0-\bar D^0$ mixing observables}
\label{sec:DDbar_observables}
%%%%%%%%%%%%%%%%%%%%%%%%%%%%%%%%%%%%%%%%%%%%%%%%%%%%%%%%%%%%%%%%%%

The $D^0-\bar D^0$ mixing amplitude can be described by means of the dispersive ($M_{12}$) 
and the absorptive ($\Gamma_{12}$) parts as follow~\cite{Bergmann:2000id,Gedalia:2009kh,
Grossman:2009mn,Bigi:2009df,Kagan:2009gb,Dighe:2013epa}
\begin{eqnarray}
\langle D^0 |\mathcal{H}_{\rm eff}| \bar D^0 \rangle &=& M_{12} - \frac{i}{2} \Gamma_{12} \, , \nn \\
\langle \bar D^0 |\mathcal{H}_{\rm eff}| D^0 \rangle &=& M_{12}^* - \frac{i}{2} \Gamma_{12}^* \, .
\label{eq:mixing_amplitude}
\end{eqnarray}
The mass eigenstates $D_{H,L}$ for the neutral $D$ meson systems are linear
combinations of the strong interaction eigenstates, $D^0$ and $\bar D^0$
\begin{equation}
| D_{L,H} \rangle = \dfrac{1}{\sqrt{|p|^2 + |q|^2}} ( \, p | D^0 \rangle \pm q | \bar D^0 
\rangle \, ) \, ,
\end{equation}
where
\begin{equation}
\label{q/p}
\frac{q}{p} =
\sqrt{\frac{M_{12}^*-\frac{i}{2}\Gamma_{12}^*}{M_{12}-\frac{i}{2}\Gamma_{12}}} \, \, .
\end{equation}
The normalized mass difference $x$ and width difference $y$ are given by
\begin{eqnarray}
x &=& \frac{\Delta M_D}{\Gamma_{D}} = \frac{M_H-M_L}{\Gamma_D} = 
2 \, \tau \, 
\textnormal{Re}\left[ \frac{q}{p} \left( M_{12} - \frac{i}{2} \Gamma_{12} \right) \right] \, , \nn \\
y &=&  \frac{\Delta \Gamma_D}{2 \Gamma_{D}} = \frac{\Gamma_H-\Gamma_L}{2\Gamma_D} = 
-2 \, \tau \, \textnormal{Im}\left[\frac{q}{p}\left( M_{12}-\frac{i}{2}\Gamma_{12}\right)\right]\,,
\end{eqnarray}
with $\tau = 1/\Gamma_{D} = 0.41$ ps \cite{pdg:2014} being the neutral $D$ life-time and $\Gamma_{D}$ 
the average decay width of the neutral $D$ mesons: $\Gamma_{D} = \frac{\Gamma_H+\Gamma_L}{2}$.
\footnote{Hereafter, $M_{12}$, $\Gamma_{12}$, $x$, $y$ and $\tau$ correspond to the $D$ system.} 
The mass difference $\Delta M_D$ is always taken to be positive by definition. 
However, the sign of $\Delta \Gamma_D$ is physically meaningful. 
Note that, our definition of $\Delta \Gamma_D$ is consistent with the HFAG convention \cite{hfag}.

In addition, we define the decay amplitudes to final state $f$ as
\begin{align}
A_f = \langle f |\mathcal{H}_{\rm eff}| D^0 \rangle \, , \qquad
\bar{A}_f = \langle f |\mathcal{H}_{\rm eff}| \bar{D}^0 \rangle \, ,
\end{align}
and the complex dimensionless parameter
\begin{align}
\lambda_f = \frac{q}{p} \frac{\bar{A}_f}{A_f} \, \, .
\end{align}
The deviation of $|q/p|$ from unity corresponds to CP violation in mixing. An example of 
this type of CP violation is the semileptonic decay asymmetry to ``wrong sign'' leptons 
$a_{\rm SL}$
\be
a_{\rm SL} = \frac{\Gamma(D^0 \to K^+ \ell^- \nu) - \Gamma(\bar D^0 \to K^- \ell^+ \nu)}{\Gamma(D^0 \to K^+ \ell^- \nu) + \Gamma(\bar D^0 \to K^- \ell^+ \nu)}  
= \frac{|q|^4-|p|^4}{|q|^4+|p|^4} \, .
\ee
When the final state $f$ is a CP eigenstate $f_{CP}$ (e.g., $\pi^+ \pi^-$, $K^+ K^-$), 
a CP violating asymmetry $A_\Gamma$ can be constructed taking the difference of the 
``effective decay width"\footnote{The ``effective decay width" is extracted by fitting the 
time-dependent decay rate to pure exponentials.} (denoted by $\hat{\Gamma}$ below)
of $D \to f_{CP}$ and  $\bar{D} \to f_{CP}$
\begin{eqnarray}
A_\Gamma(f_{CP}) &=&
\frac{\hat\Gamma_{D^0 \to f_{CP}} - \hat\Gamma_{\bar D^0 \to f_{CP}}}{\hat\Gamma_{D^0 \to f_{CP}} + \hat\Gamma_{\bar D^0 \to f_{CP}}} 
\label{eq:DYf} \nn \\ 
& \simeq& \frac{y}{2} \bigg[{\rm Re}(\lambda_{f_{CP}}) - {\rm Re}(\lambda^{-1}_{f_{CP}})\bigg] -\frac{x}{2} \bigg[{\rm Im}(\lambda_{f_{CP}}) - {\rm Im}(\lambda^{-1}_{f_{CP}})\bigg]  
\nn \\
& = & \frac{y}{2}\left(\left|\frac{q}{p}\right|-\left|\frac{p}{q}\right|\right)\cos\phi - \frac{x}{2}\left(\left|\frac{q}{p}\right|+\left|\frac{p}{q}\right|\right)\sin\phi\nn \\
&\approx& \frac{y}{2}\left(\left|\frac{q}{p}\right|-\left|\frac{p}{q}\right|\right)- x\sin\phi \,.
\end{eqnarray}
The above expression has been obtained assuming $\bar{A}_f/A_f = 1$ and working to linear order in the CP violating parameters. 
In the absence of direct CP violation $A_\Gamma$ and $a_{\rm SL}$ (or $\sin\phi$) are correlated by the model-independent 
relation~\cite{Ligeti:2006pm,Ciuchini:2007cw,Grossman:2009mn} 
\bea
A_\Gamma  &\approx& \dfrac{x^2 + y^2}{y} ~ \dfrac{a_{\rm SL}}{2} = x\sin \Phi_{12} \approx - \frac{x^2 + y^2 }{x}\sin \phi \,.
\label{eq:agamma_asl}
\eea
As far as the experimental situation is concerned, the most recent fit results from the UTfit 
collaboration are collected in Tab.~\ref{tab:ddmix_res}.
\begin{table}[t]
  \centering
  \begin{tabular}{|ccc|}
    \hline
    parameter & result @ $68\%$ prob. & $95\%$ prob. range\\
    \hline
    \hline
    $|M_{12}|$[ps$^{-1}]$ & $(4.4\pm 2.0) \cdot 10^{-3}$ & $[0.3,7.7]\cdot 10^{-3}$ \\
    $|\Gamma_{12}|$[ps$^{-1}]$ & $(14.9\pm 1.6) \cdot 10^{-3}$ & $[11.7,18.5]\cdot 10^{-3}$ \\     
    $\Phi_{{12}}[\deg]$ & $(2.0\pm 2.7)$ & $[-4,12]$ \\
    \hline
    $x$ & $(3.6\pm 1.6) \cdot 10^{-3}$ & $[0.3,6.7]\cdot 10^{-3}$ \\
    $y$ & $(6.1\pm 0.7) \cdot 10^{-3}$ & $[4.8,7.6]\cdot 10^{-3}$ \\
    $\vert q/p\vert$ & $1.016\pm 0.018$ & $[0.981,1.058]$ \\
    $\phi [^\circ]$ & $-0.5\pm 0.6$ & $[-1.8,0.6]$\\
    $A_\Gamma$ & $(1.4\pm 1.5) \cdot 10^{-4}$ & $[-1.5, 4.4]\cdot 10^{-4}$ \\
    $a_{\rm\mysmall SL}$ & $(3.2\pm 3.6) \cdot 10^{-2}$ & $[-3.8, 11.3]\cdot 10^{-2}$ \\
%    $y_\mathrm{\mysmall CP}$ & $(6.1\pm 0.7) \cdot 10^{-3}$ & $[4.8, 7.6]\cdot 10^{-3}$ \\
  \hline
  \end{tabular}
  \caption{Results of the fit to $D$ mixing data from the UTfit collaboration~\cite{UTfit_D}.}
  \label{tab:ddmix_res}
\end{table}
Even if $D^0-\bar D^0$ mixing is now firmly established experimentally, there is no evidence yet for CP violation. 
In particular, current data are compatible with the hypothesis of CP conservation, i.e. $|q/p|=1$ and $\phi=0$ to a 
better than 10\% accuracy. This justifies our linear expansion of CP violating quantities.

Eq.~\eqref{eq:agamma_asl} can be further used to constrain the phase of a heavy NP.
We shall assume here that the SM contributions are dominated by the first two generations and thus can be brought to be real without loss of generality.
Thus, any CP violation can only arise due to the presence of an imaginary component of the dispersive part of the $\Delta c=2$ amplitude, $\IM$,
\be
a_{\rm SL} \simeq 2 \left( \left|\frac{q}{p}\right| - 1 \right) \simeq \frac{y}{x^2+y^2} \, 4 \tau \IM\,,\label{IM}
\ee
see~\cite{Kagan:2009gb,Gedalia:2009kh} for more details.

NP effects for $D^0-\bar D^0$ mixing can be described in full generality
by means of the $\Delta F = 2$ effective Hamiltonian
\begin{equation}
\mathcal{H}_{\rm eff} = \sum_{i=1}^5 C_i Q_i +
\sum_{i=1}^3 \tilde C_i \tilde Q_i ~+{\rm h.c.}~,
\label{eq:WCs}
\end{equation}
where $C_i$ are the Wilson Coefficients (WCs) of the operators $Q_i$ given by
\begin{eqnarray}
Q_1 & = &
(\bar u^\alpha \gamma_\mu P_L c^\alpha)(\bar u^\beta \gamma^\mu P_L c^\beta)~,\nonumber \\
Q_2 & = &
(\bar u^\alpha P_L c^\alpha)(\bar u^\beta P_L c^\beta)~,\nonumber \\
Q_3 & = &
(\bar u^\alpha P_L c^\beta)(\bar u^\beta P_L c^\alpha)~,\nonumber \\
Q_4 & = &
(\bar u^\alpha P_L c^\alpha)(\bar u^\beta P_R c^\beta)~,\nonumber \\
Q_5 & = &
(\bar u^\alpha P_L c^\beta)(\bar u^\beta P_R c^\alpha)~,
\label{eq:DF2_operators}
\end{eqnarray}
where $P_{R,L}=\frac{1}{2}(1\pm\gamma_5)$ and $\alpha,\beta$ are colour indices.
The operators $\tilde{Q}_{1,2,3}$, which we have omitted, are obtained from
$Q_{1,2,3}$ through the replacement $L \leftrightarrow R$.

For the calculation of the observables, we have used the hadronic matrix elements
and the magic numbers from \cite{Bona:2007vi}.

%%%%%%%%%%%%%%%%%%%%
\section{Model-independent analysis} 
\label{sec:mod_ind_analysis}
In general, NP effects for $\Delta F = 1,2$ transitions in the up- and down-quark sectors are unrelated. 
As such, the very stringent constraints arising from FCNC processes like $\epsilon^\prime/\epsilon$ or 
$\epsilon_K$ do not necessarely imply similar constraints on FCNC processes involving $D$ mesons. 
Yet, there are many NP scenarios in which the dominant effects are encoded in operators involving only 
the quark-doublet $q_L$. 
In such cases, FCNC contributions for $K$ and $D$ meson systems stem from the fermionic bilinear $\overline{q_{L}} \gamma_\mu q_{L}$
and therefore are approximately $SU(2)_L$ invariant~\cite{Blum:2009sk}. 

Focusing on these scenarios, the relevant $\Delta F = 1,2$ operators are, respectively
\begin{align}
&\frac{1}{\Lambda_{\rm NP}^2}\left(\overline{q_{Li}} \, X_{ij} \, \gamma_\mu q_{Lj} \right) O^\mu   \qquad\qquad\qquad\qquad\qquad  \Delta F = 1~,
\\
&\frac{1}{\Lambda_{\rm NP}^2}\left(\overline{q_{Li}} \, X_{ij} \, \gamma_\mu q_{Lj} \right) \left(\overline{q_{Li}} \, X_{ij} \, \gamma^\mu q_{Lj}\right) ~~\qquad\qquad~ 
\Delta F = 2~,
\end{align}
where $O^\mu = \sum_q \overline{q} \gamma^\mu q$,  $\sum_\ell \overline{\ell} \gamma^\mu \ell$, etc.
Since $X$ is an hermitian matrix it can be diagonalised through a unitary matrix $V$ as $X=V^\dagger {\hat X} V$ 
where ${\hat X} = {\rm diag}({\hat X}_1,{\hat X}_2)$ and ${\hat X}_{1,2}$ are the eigenvalues of $X$. In the down mass 
basis it turns out that 
\begin{align}
& \frac{1}{\Lambda_{\rm NP}^2} \left[c_K (\overline{d_L}\gamma_\mu s_L) + c_D (\overline{u_L}\gamma_\mu c_L) \right] O^\mu ,\\
&\frac{1}{\Lambda_{\rm NP}^2} \left[z_K (\overline{d_L}\gamma_\mu s_L)(\overline{d_L}\gamma^\mu s_L) +
z_D (\overline{u_L}\gamma_\mu c_L)(\overline{u_L}\gamma^\mu c_L)\right],
\end{align}
where $c_K (z_K)$ and $c_D (z_D)$ are related through the CKM matrix $V_{\mysmall {\rm CKM}}$ as follow
\begin{align}
%z_K = c^2_K = \left[ (V_q^\dagger {\hat X}_q V_q)_{12} \right]^2, \qquad
%z_D = c^2_D = \left[( V_{\mysmall {\rm CKM}} V_q^\dagger {\hat X}_q V_q V_{\mysmall {\rm CKM}}^\dagger)_{12} \right]^2.
z_K = c^2_K = \left( {X}_{12} \right)^2, \qquad
z_D = c^2_D = \left[( V_{\mysmall {\rm CKM}} X V_{\mysmall {\rm CKM}}^\dagger)_{12} \right]^2.
\end{align}
Working in a two-generation framework, which is appropriate for our purposes,  $V_{\mysmall {\rm CKM}}$ and $V$
can be parametrised as follows
\be
V_{\mysmall {\rm CKM}} = 
\left(\begin{array}{cc}
\cos\theta_{\rm C} & \sin\theta_{\rm C} \\ -\sin\theta_{\rm C} & \cos\theta_{\rm C}
\end{array}\right), \qquad
V = \left(\begin{array}{cc}
\cos\theta_q & \sin\theta_q \, e^{i\alpha}\\ -\sin\theta_q \, e^{-i\alpha}& \cos\theta_q
\end{array}\right).
\label{eq:V_VCKM}
\ee
As a result, the coefficients $c_{K(D)}$ governing $\Delta F =1$ transitions read
\begin{align}
{\rm Re} \,c_K &= \frac{\sin 2\theta_q}{2} ( {\hat X}_1 - {\hat X}_2 ) \cos\alpha\,,
\\
{\rm Im} \,c_K &= {\rm Im} \,c_D = \frac{\sin 2\theta_q}{2} ( {\hat X}_1 - {\hat X}_2 ) \sin\alpha\,,
\\
{\rm Re} \,c_D &= \frac{( {\hat X}_1 - {\hat X}_2 ) }{2} \left(\cos 2\theta_{\rm C} \sin 2\theta_q \cos\alpha  - \cos 2\theta_q \sin 2\theta_{\rm C}  \right)\,.
\end{align}
In particular, the relation ${\rm Im} \,c_K = {\rm Im} \,c_D$ implies that, within our framework,
CP violating effects in $\Delta F = 1$ transitions are universal in the up- and down-quark sectors,
in agreement with~\cite{Gedalia:2012pi}.

Passing to $\Delta F = 2$ transitions, we find the following results 
\begin{align}
{\rm Re} \,z_K &= \frac{\sin^2 2\theta_q}{4} ( {\hat X}_1 - {\hat X}_2 )^2 \cos 2\alpha\,,
\\
{\rm Im} \,z_K &= \frac{\sin^2 2\theta_q}{4} ( {\hat X}_1 - {\hat X}_2 )^2 \sin 2\alpha\,,
\\
{\rm Im} \,z_D &= \frac{\sin 2\theta_q}{2} ( {\hat X}_1 - {\hat X}_2 )^2 \left( \cos 2\theta_{\rm C} \sin 2\theta_q  \cos\alpha  - \sin 2\theta_{\rm C} \cos 2\theta_q \right) \sin\alpha\,,
\\
{\rm Re} \,z_D &= \frac{( {\hat X}_1 - {\hat X}_2 )^2}{16} 
\bigg[(1 + 3 \cos 4\theta_q) \sin^2 2\theta_{\rm C}  - 2 \sin 4\theta_{\rm C} \sin 4\theta_q \cos\alpha +
\nonumber\\
& \qquad\qquad\qquad~~~
(3 + \cos 4\theta_c) \sin^2 2\theta_q \cos 2 \alpha
\bigg]\,. 
\end{align}
Let us simplify the above expressions remembering that $\cos\theta_{\rm C} \approx 1$, $\sin\theta_{\rm C} \approx \lambda_{\rm C} \approx 0.22$ and taking the limit of almost 
alignment where $\theta_q \ll 1$. We find that
\begin{align}
{\rm Re} \,c_K &= ( {\hat X}_1 - {\hat X}_2 )\, \theta_q \cos\alpha\,,\nn
\\
{\rm Im} \,c_K &= {\rm Im} \,c_D = ( {\hat X}_1 - {\hat X}_2 )\, \theta_q \sin\alpha\,,\nn
\\
{\rm Re} \,c_D &= ( {\hat X}_1 - {\hat X}_2 ) \left(\theta_q \cos\alpha  - \lambda_{\rm C}  \right)\,,\nn
\\
{\rm Re} \,z_K &= ( {\hat X}_1 - {\hat X}_2 )^2 \, \theta^2_q \cos 2\alpha\,,\nn
\\
{\rm Im} \,z_K &= ( {\hat X}_1 - {\hat X}_2 )^2  \, \theta^2_q  \sin 2\alpha\,,\nn
\\
{\rm Im} \,z_D &= ( {\hat X}_1 - {\hat X}_2 )^2  
( \theta^2_q  \sin 2\alpha  - 2\lambda_{\rm C}\, \theta_q \sin\alpha \nn
) \,,
\\
{\rm Re} \,z_D &= ( {\hat X}_1 - {\hat X}_2 )^2 
(
\lambda^2_{\rm C}  + \theta^2_q \cos 2 \alpha - 2 \lambda_{\rm C} \,\theta_q \cos\alpha
) \,.
\label{eq:approx_expressions}
\end{align}
\begin{figure*}[t!]
\begin{center}
\includegraphics[width=0.45\textwidth]{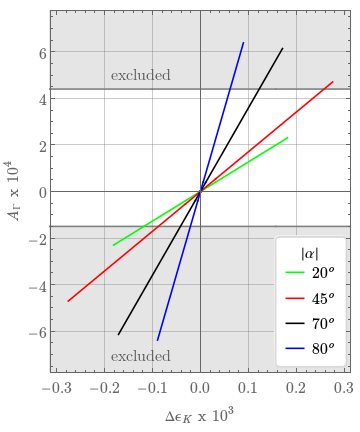}~
\includegraphics[width=0.45\textwidth]{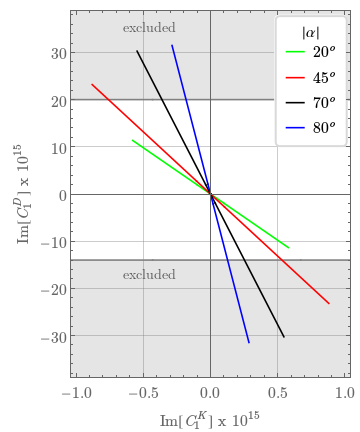}
\caption{
Left: correlation between $A_{\Gamma}$ and $\Delta \epsilon_K$.
Right: correlation between ${\rm Im} \left(C^D_1\right)$ and ${\rm Im} \left(C^K_1\right)$. 
The plots have been obtained imposing the bound on $|M_{12}|$ of Tab~\ref{tab:ddmix_res}.
In the 1st and 4th quadrant $\alpha = (20°, 45°, 70°, 80°)$ while in the 2nd and 3rd quadrant 
$\alpha = -(20°, 45°, 70°, 80°)$.}
\label{fig:model_indep_plots}
\end{center}
\end{figure*}
The expressions above show that CP violating effects entering $K$ and $D$ meson systems are not universal 
for $\Delta F = 2$ transitions.
Yet, it is still possible to obtain a model-independent upper bound for charm CP violating effects. 
In order to see this, we notice that that above relations imply
\begin{align}
|{\rm Im} \,z_D| 
%&
\approx  \sqrt{2 \tan\alpha \times {\rm Im} \,z_K} ~ \sqrt{{\rm Re} \, z_D} \, 
%& 
~\lesssim ~4 \times 10^{-8} \, \sqrt{|\tan\alpha|}
\,,
\label{eq:mod_ind_bound}
\end{align}
where the upper bound on $|{\rm Im} \,z_D|$ has been obtained assuming the bounds on $|{\rm Im} \,z_K|$ and 
$|{\rm Re} \,z_D|$ from refs.\cite{Bertone:2012cu} and \cite{Carrasco:2014uya}, respectively. 
Since we are interested in a relation among physical observables, we exploit the model-independent results of the 
previous section in the limit of small CP violation. In particular, from Eq.~\eqref{eq:agamma_asl} and~\eqref{IM}, 
and remembering that in D-physics we are interested in the two generation limit, where all the SM couplings can be made real 
without loss of generality,  we have
\begin{equation}
A_\Gamma/2\tau \sim \IM_{\rm SM\in Real} \propto{\rm Im} \, z_D
\label{AgamNP} \,.
\end{equation}
As a result, it is straightforward to find the following expression for $|A_\Gamma|$
\begin{align}
|A_\Gamma|
\lesssim 0.36 ~  \sqrt{x \, \tan\alpha ~ \Delta\epsilon_K} \,,
\label{eq:mod_ind_bound_obs}
\end{align}
where $\Delta \epsilon_K \sim {\rm Im} \, z_K$.
Finally, imposing the experimental bounds on $x$ and $\Delta\epsilon_K$, we can find the desired theoretical upper bound for $|A_\Gamma|$
\begin{align}
|A_\Gamma| 
\lesssim  9.3 \times 10^{-4}  \,  \sqrt{\frac{|\Delta\epsilon_K|^{\rm max}}{1.0 \times 10^{-3}}} \, \sqrt{\frac{x^{\rm max}}{6.7 \times 10^{-3}}} \, \sqrt{|\tan\alpha|} \,,
\label{eq:mod_ind_bound_obs2}
\end{align}
where we have assumed $|\Delta\epsilon_K|^{\rm max} \lesssim 1.0 \times 10^{-3}$ such that $|\Delta\epsilon_K|^{\rm max}/\epsilon_K^{\rm\mysmall SM} \lesssim 0.4$\,,
in agreement with the bound quoted in ref.~\cite{Bertone:2012cu}.
Therefore, the current experimental resolutions (see Tab.~1) are testing right now the natural predictions of alignment models. 

In Fig.\,\ref{fig:model_indep_plots} on the left, we show the model-independent correlation between $A_{\Gamma}$ and $\Delta \epsilon_K$ within alignment models. 
As we can explicitly see, positive NP effects for $\Delta \epsilon_K$ at the level of $20\%-30\%$ (which would even improve the current UTfit analyses) naturally 
imply values for $A_\Gamma$ close to the present bound  $A_\Gamma \lesssim 4.4\times 10^{-4}$. In fig.\,\ref{fig:model_indep_plots} on the right we show also 
the correlation between ${\rm Im} \left(C^D_1\right)$ and ${\rm Im} \left(C^K_1\right)$. In both plots we have imposed the bound on $|M_{12}|$ of Tab~\ref{tab:ddmix_res} 
and set $\theta_q = \lambda_{\rm C}^3$ for definiteness. 
Moreover, we have considered $\alpha = (20°, 45°, 70°, 80°)$ in the 1st and 4th quadrants while $\alpha = -(20°, 45°, 70°, 80°)$ in the 2nd and 3rd quadrants.

%%%%%%%%%%%%%%%%%%%%%%%%%%%%%%%%%%%%%%%%%%%%%%%%%%%%%%%%%%%%%%%%%%%%%%%%%%%%%%
\section{$D^0-\bar D^0$ mixing in SUSY Alignment Models} 
\label{sec:DDbar_align}
%%%%%%%%%%%%%%%%%%%%%%%%%%%%%%%%%%%%%%%%%%%%%%%%%%%%%%%%%%%%%%%%%%%%%%%%%%%%%%

We shall now move to consider SUSY alignment models~\cite{Nir:1993mx,Leurer:1993gy}.
It amounts to aligning the squark and quark mass matrices either in the up- or down-sector, 
so that FCNC effects are kept under control without requiring any degeneracy in the squark 
spectrum.

As argued in Ref.~\cite{Nir:2002ah}, within alignment models it is possible to predict both lower and 
upper bounds for the SUSY flavor mixing angles $(s^q_{\mysmall M})_{ij}$ entering the couplings 
$\tilde g{\mysmall -}q_{{\mysmall M}_i} {\mysmall -} \tilde q_{{\mysmall M}_j}$, with $M = L, R$. 
In particular, by making use of holomorphic zeros in the down quark mass matrix to suppress the 
mixing angles of the first two generations, one can find the predictions of Tab.~\ref{QSA}.
%
%%%%%%%%%%%%%%%%%%%%%%%%%%%%%%%%%%%%%%%%%%%%%%%%%%%%%%%%%%%%%%%%%%
\begin{table}
\centering
\begin{tabular}{c|c|c}
\hline\hline
Mixing Angle  &  Lower Bound  & Upper Bound  \\
\hline\hline
$(s^d_{L})_{12}$ & $\lambda_{\rm C}^5$ & $\lambda_{\rm C}^3$ \\
$(s^d_{R})_{12}$ & $\lambda_{\rm C}^7$ & $\lambda_{\rm C}^3$ \\
\hline
$(s^d_{L})_{13}$ & $\lambda_{\rm C}^3$ & $\lambda_{\rm C}^3$ \\
$(s^d_{R})_{13}$ & $\lambda_{\rm C}^7$ & $\lambda_{\rm C}^3$ \\
\hline
$(s^d_{L})_{23}$ & $\lambda_{\rm C}^2$ & $\lambda_{\rm C}^2$ \\
$(s^d_{R})_{23}$ & $\lambda_{\rm C}^4$ & $\lambda_{\rm C}^2$ \\
\hline
$(s^u_{L})_{12}$ & $\lambda_{\rm C}$   & $\lambda_{\rm C}$ \\
$(s^u_{R})_{12}$ & $\lambda_{\rm C}^4$ & $\lambda_{\rm C}^2$ \\
\hline\hline
\end{tabular}\vspace*{4pt}
\caption{Lower and upper bounds on SUSY flavor mixing angles in alignment models~\cite{Nir:2002ah}.}
\label{QSA}
\end{table}
%%%%%%%%%%%%%%%%%%%%%%%%%%%%%%%%%%%%%%%%%%%%%%%%%%%%%%%%%%%%%%%%%%

The most prominent feature of these models is the appearance of a large left-handed 
mixing between the first two families. In particular, in the so-called super-CKM basis, 
the left-handed squark mass matrices are related by the $SU(2)_L$ relation 
$M^{2}_{{\tilde u}, LL}=V_{\rm CKM} \,  M^{2}_{{\tilde d}, LL} V_{\rm CKM}^{\dagger}$, 
where $V_{\rm CKM}$ is the CKM matrix. 
A leading order expansion in the Cabibbo angle leads to the following expression 
\be
\label{qsa_approx}
(M^{2}_{{\tilde u}, LL})_{21} =
(M^{2}_{{\tilde d}, LL})_{21} + \lambda_{\rm C} \left[(M^{2}_{{\tilde d}, LL})_{22}-(M^{2}_{{\tilde d}, LL})_{11}\right] + \mathcal{O}(\lambda_{\rm C}^2) \, . 
\ee
Therefore, even assuming a perfect alignment in the down sector, that is $(M^{2}_{{\tilde d}, LL})_{21}=0$, 
we always end up with a large flavor violating entry in $(M^{2}_{{\tilde u}, LL})_{21}$ proportional to $\lambda_{\rm C}$ 
as long as the left-handed squarks are non-degenerate.

The usual prescription is to start from Eq.~(\ref{qsa_approx}) and define the following MI~\cite{Raz:2002zx}
\be
(\delta_{u}^{L})_{21} =
(\delta_{d}^{L})_{21} + \lambda_{\rm C} ~ 
\frac{(M^{2}_{{\tilde d}, LL})_{22}-(M^{2}_{{\tilde d}, LL})_{11}}{m_{\tilde q}^2} 
=
(\delta_{d}^{L})_{21} + 4\, \lambda_{\rm C} \, \xi\,,
\label{defdel}
\ee
where, considering only the first two generations, $(\delta_{u,d}^{L})_{21}$, $\xi$ and $m_{\tilde q}$ read
\begin{align}
(\delta_{u}^{L})_{21} &= \frac{(M^{2}_{{\tilde u}, LL})_{21}}{m_{\tilde q}^2}\,, \qquad
(\delta_{d}^{L})_{21} = \frac{(M^{2}_{{\tilde d}, LL})_{21}}{m_{\tilde q}^2} \, ,  \\
m_{\tilde q} &= \frac{\sqrt{(M^2_{\tilde Q})_{11}} + \sqrt{(M^2_{\tilde Q})_{22}}}{2} \, , \\
\xi &= \frac{(M^{2}_{{\tilde d}, LL})_{22}-(M^{2}_{{\tilde d}, LL})_{11}}
{\bigg(\sqrt{(M^2_{\tilde Q})_{11}} + \sqrt{(M^2_{\tilde Q})_{22}}\bigg)^2} 
\label{eq:xi}\, .
\end{align}
Here, $M^2_{\tilde Q}$ is the squark mass matrix squared for the left-handed squark-doublets. 
As a result, flavor constraints translate into constraints on SUSY masses and the mass splitting parameter $\xi$.
If the mass splittings among squarks is sizable, i.e. $\xi \sim 1$, the MI approximation is not in general a good 
approximation, as we will discuss quantitatively in the following.

The main goal of the following section is twofold:
\begin{itemize}
\item to derive exact analytical expressions for $C_i$, see Eq.~\eqref{eq:WCs},
working in a two-generation framework and performing an analytical diagonalization 
of the squark mass matrices.
We account for the full set of SUSY contributions which include pure gluino, mixed
neutralino/gluino, chargino, as well as neutralino effects;~\footnote{The gluino
contributions to $C_i$ have been already evaluated in the MI approximation at the
LO in~\cite{Gabbiani:1996hi} and at NLO in~\cite{Ciuchini:2006dw}. The full set of 
LO contributions in the mass-eigenstate basis and with three generations has been 
presented in ref.~\cite{Altmannshofer:2007cs}. Although our results can be regarded 
as a special case of those of ref.~\cite{Altmannshofer:2007cs}, we stress that our 
expressions for $C_i$ have the major advantage of being much simpler (as they do 
not require any numerical diagonalization procedure) while reproducing the numerical 
results based on ref.~\cite{Altmannshofer:2007cs} with an excellent accuracy.
}
\item to derive the expressions for $C_i$ in the MI approximation in two relevant limits for 
the squark masses: the quasi-degeneracy and split scenarios, clarifying the extent to which 
the commonly used MI approximation (so far known only for the gluino contributions) agrees 
with the exact computation.
\end{itemize}

%%%%%%%%%%%%%%%%%%%%%%%%%%%%%%%%%%%%%%%%%%%%%%%%%%%%%%%%%%%%%%%%%%%
\subsection{Full results}
\label{sec:full_results}
%%%%%%%%%%%%%%%%%%%%%%%%%%%%%%%%%%%%%%%%%%%%%%%%%%%%%%%%%%%%%%%%%%%
%
In the following, we provide the relevant expressions for $C_i$ and ${\tilde C}_i$
in SUSY alignment models under the following approximations:
\begin{enumerate}
\item we work in a two-generation framework. Such an approximation is justified
if the underlying $c \to u$ transition is not significantly affected by flavor
mixings with the third generation, that is if the direct $c \to u$ transition
dominates over the double flavor transition $(c \to t) \times (t \to u)$. 
This is an excellent approximation in alignment models, as one can check from 
Tab.~\ref{QSA};
\item we neglect the small Yukawa couplings for the first two generations and therefore
the corresponding LR/RL soft terms while we keep the full dependence on the chargino
and neutralino mixings;
\item we neglect $U(1)_Y$ interactions since they are safely negligible compared to 
$SU(2)_L$ interactions, as we have explicitly checked numerically.
\end{enumerate}
The most important effects for $D^0-\bar D^0$ mixing in alignment models arise
from the operators $Q_1$ and $Q_{4,5}$ since their (different) sensitivity to the 
large MI $(\delta_{u}^{L})_{21}\sim\lambda_{\rm C}$. Our results for $C_1$ are given by 
the following expressions
\begin{align}
C^{\g\g}_1 &=
-\frac{\alpha^2_s}{9} \left(s_{\mysmall L}c_{\mysmall L}\,e^{i\phi_L}\right)^2 
\bigg( m_\g^2 \, B_{\tilde{u}_L\tilde{u}_L}(m_\g^2, m_\g^2) + 11 \, C_{\tilde{u}_L\tilde{u}_L}(m_\g^2, m_\g^2) \bigg) \, , 
\label{eq:C1gg}
\\
C^{\chi^+\!\chi^+}_1 &=
-\frac{\alpha^2_w}{2}
\left(s_{\mysmall L}c_{\mysmall L}\,e^{i\phi_L}\right)^2 (Z^{1a}_{-})^2 \, (Z^{1b}_{-})^2 
C_{\tilde{d}_L\tilde{d}_L}\left(M^{2}_{\chi^{\pm}_{a}}, M^{2}_{\chi^{\pm}_{b}}\right) \, ,
\label{eq:C1chichi}
\\
C_1^{\chi^0\!\chi^0} &=
- \frac{\alpha^2_w}{8}\left(s_{\mysmall L}c_{\mysmall L}\,e^{i\phi_L}\right)^2 \left(Z^{2a}_{N} \right)^2 \! \left(Z^{2b}_{N} \right)^2 \times
\nn \\
& ~~~\bigg(\frac{1}{2} M_{\chi^{0}_{a}}M_{\chi^{0}_{b}}~B_{\tilde{u}_L\tilde{u}_L}
(M^{2}_{\chi^{0}_{a}}, M^{2}_{\chi^{0}_{b}})
+ C_{\tilde{u}_L\tilde{u}_L}(M^{2}_{\chi^{0}_{a}}, M^{2}_{\chi^{0}_{b}})
\bigg)\, ,
\label{eq:C1chi0chi0}
\\
C^{\g \chi^0}_1 &=
-\frac{\alpha_s \alpha_w}{3} \left(s_{\mysmall L}c_{\mysmall L}\,e^{i\phi_L}\right)^2 
\left(Z^{2a}_{N} \right)^2\times
\nn \\
& ~~~
\bigg( \frac{1}{2} M_{\chi^{0}_{a}} m_\g~B_{\tilde{u}_L\tilde{u}_L}(M^{2}_{\chi^{0}_{a}},m^2_\g) +
C_{\tilde{u}_L\tilde{u}_L}(M^{2}_{\chi^{0}_{a}},m^2_\g) \bigg) \,,
\label{eq:C1gchi0}
\end{align}
where a sum over the indices $a,b=1,4$ (for neutralinos) and $a,b=1,2$
(for charginos) is undesrtood. The matrices $Z_{N}$ and $Z_{-}$, which
stem from the diagonalization of the chargino and neutralino mass matrices,
as well as the mixing angles $s_{\mysmall L}c_{\mysmall L}\,e^{i\phi_L}$
are defined in Appendix A while the loop functions $B(x,y)$ and $C(x,y)$ 
are defined in Appendix B. The WCs $C^{gg}_1$, $C^{\chi^+\chi^+}_1$, 
$C_1^{\chi^0\chi^0}$, $C^{\tilde g \chi^0}_1$ stand for the pure gluino,
chargino, neutralino, and mixed neutralino/gluino effects, respectively.

If in addition to left-handed mixings right-handed mixings might also be present, 
thus for completeness we present the relevant functions, ${\tilde C}_{1}$ and $C_{4,5}$ 
\begin{align}
{\tilde C}^{\g\g}_1 &=
-\frac{\alpha^2_s}{9} \left(s_{\mysmall R}c_{\mysmall R}\,e^{i\phi_R}\right)^2 
\bigg( m_\g^2~B_{\tilde{u}_R\tilde{u}_R}(m_\g^2, m_\g^2) + 
11 \, C_{\tilde{u}_R\tilde{u}_R}(m_\g^2, m_\g^2) \bigg)~,
\label{eq:C1tildegg}
\\ 
C^{\g\g}_4 &=
-\frac{\alpha^2_s}{3}
\left(s_{\mysmall L}c_{\mysmall L}e^{i\phi_{\mysmall L}}\right)
\left(s_{\mysmall R}c_{\mysmall R}e^{i\phi_{\mysmall R}}\right)
\bigg( 7 m_\g^2~B_{\tilde{u}_L\tilde{u}_R}(m_\g^2, m_\g^2) - 
4 \, C_{\tilde{u}_L\tilde{u}_R}(m_\g^2, m_\g^2) \bigg)~,
\label{eq:C4gg}
\\
C^{\g\g}_5 &=
-\frac{\alpha^2_s}{9}
\left(s_{\mysmall L}c_{\mysmall L}e^{i\phi_{\mysmall L}}\right)
\left(s_{\mysmall R}c_{\mysmall R}e^{i\phi_{\mysmall R}}\right) 
\bigg( m_\g^2~B_{\tilde{u}_L\tilde{u}_R}(m_\g^2, m_\g^2) + 
20 \, C_{\tilde{u}_L\tilde{u}_R}(m_\g^2, m_\g^2) \bigg)~.
\label{eq:C5gg}
\end{align}
Notice that ${C}_{2,3}$ and ${\tilde C}_{2,3}$ are vanishing in the limit of 
vanishing LR/RL flavor mixings, which we assume.

Few comments are in order:
\begin{itemize}
\item $C^{\g\g}_1$, $C_1^{\chi^0\chi^0}$, and $C^{\tilde g \chi^0}_1$ receive two
contributions, corresponding to crossed and uncrossed gluino and/or neutralino
lines, as a result of the Majorana nature of the gluino and neutralinos. Such
contributions have opposite sign and therefore tend to cancel to each other,
the extent of cancellations depending on the parameter space. By contrast,
$C^{\chi^+\chi^+}_1$ is not affected by any cancellation since charginos are Dirac
particles and therefore there are no crossed diagrams for $C^{\chi^+\chi^+}_1$.
\item Even if $C^{\chi^+\chi^+}_1$ is parametrically suppressed compared to $C^{\g\g}_1$ by
a factor of $\alpha^2_w/\alpha^2_s \approx 1/10$, it might still provide important/dominant
effects whenever the gluino is sufficiently heavier than squarks and charginos or when
the above cancellations in $C^{\g\g}_1$ are significant. Similar comments apply also to the
case of $C_1^{\chi^0\chi^0}$, as long as we are far from the cancellation regions 
for $C_1^{\chi^0\chi^0}$. Finally, $C^{\tilde g \chi^0}_1$ can also provide significant effects
especially when $C^{\g\g}_1$ (but not $C^{\tilde g \chi^0}_1$) is suppressed by large cancellations.
\footnote{The relevance of electroweak effects has been pointed out first in ref.~\cite{Crivellin:2010ys}}
\item Assuming the upper and lower bounds for the flavor mixing angles of Tab.~\ref{QSA},
we find that $C_{1} \propto (\xi\lambda_{\rm C})^{2}$ while $C_{4,5}\propto \xi\lambda_{\rm C}^{3-5}$ and
therefore $\xi/\lambda_{\rm C} \lesssim|C_{1}|/|C_{4,5}|\lesssim \xi/\lambda_{\rm C}^{3}$. 
Taking into account that $Q_{4,5}$ have larger hadronic matrix elements than $Q_{1}$ and that 
QCD runnings further enhance $C_{4,5}$ with respect to $C_{1}$, it turns out that the contributions 
of $C_{4,5}$ to the $D^0-\bar D^0$ mixing amplitude are very important even for $\xi\sim {\mathcal O}(1)$.
\item In the limit of complete alignment, i.e. for $(M^{2}_{{\tilde d}, LL})_{21} = (M^{2}_{{\tilde u}, RR})_{21} = 0$,
CPV effects in $D^0-\bar D^0$ mixing are vanishing~\cite{Altmannshofer:2010ad}. Possible CPV sources can 
arise only in the presence of a misalignment either in the LL or RR sectors. In the former case, the underlying 
$SU(2)_L$ symmetry links CPV effects in $D$- and $K$-meson systems. In the latter case, the above CPV 
effects are generally unrelated.
\item Naively, one would expect that flavor violating sources in the LL up-squark sector are felt by the 
down sector through chargino up-squark contributions. However, the chargino amplitude is such that
$A^{\tilde\chi}_{ij}\sim(V^{\dagger} M^{2}_{{\tilde u}, LL} V)_{ij} \equiv 
(M^{2}_{{\tilde d}, LL})_{ij}$ and therefore down-quark FCNCs turn out to be sensitive to $M^{2}_{{\tilde d}, LL}$
and not $M^{2}_{{\tilde u}, LL}$~\cite{Altmannshofer:2010ad}.
\end{itemize}
For concreteness and also to simplify the numerics, we temporarily switch-off the phases and mixing 
relative to the down-mass basis, when showing our results, namely we assume complete alignment, 
i.e. $(M^{2}_{{\tilde d}, LL})_{21} =0$ and $(M^{2}_{{\tilde u}, RR})_{21}= 0$. 

\begin{figure*}[t!]
\begin{center}
\includegraphics[scale=1]{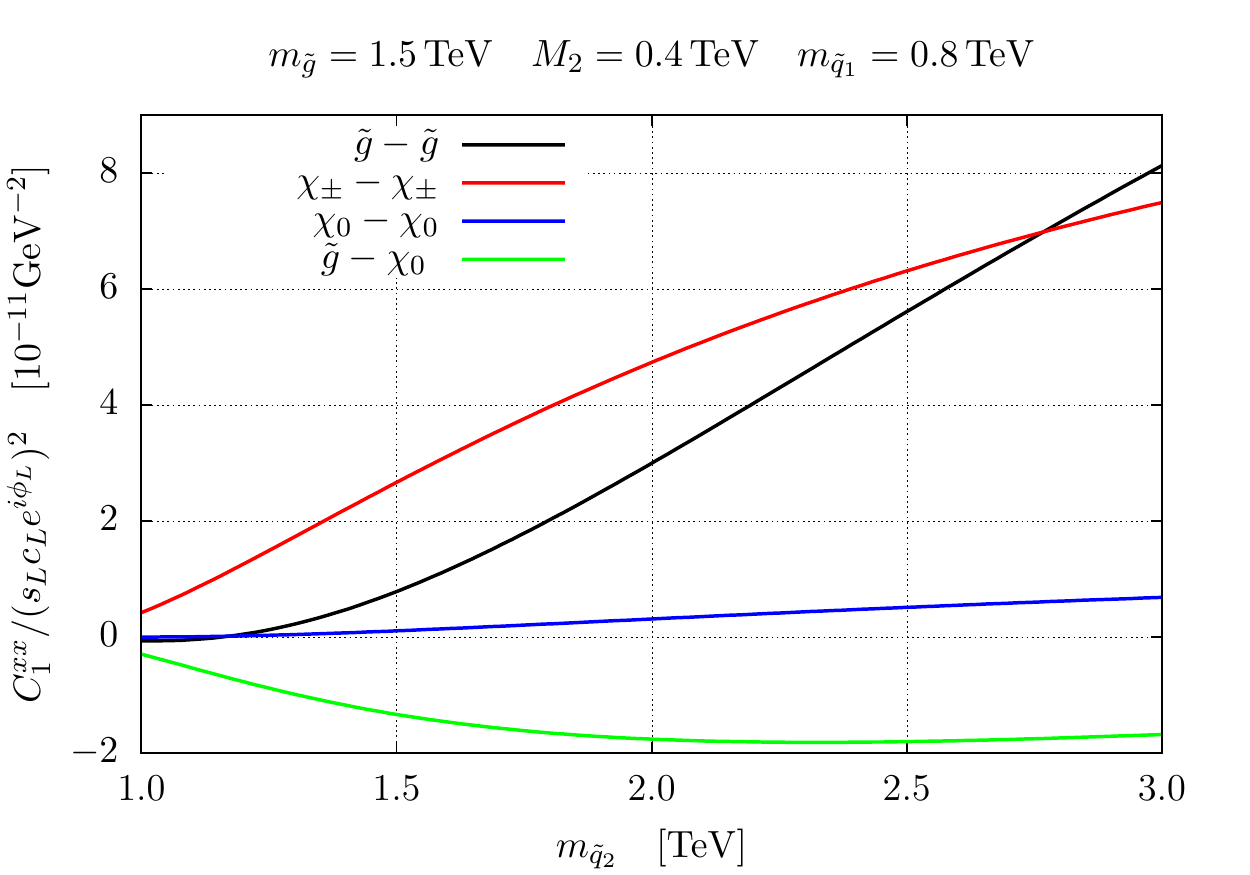}
\caption{Contributions to the Wilson coefficient $C^{xx}_1$, i.e.  $C^{\g\g}_1$, $C^{\chi^+\chi^+}_1$, 
$C_1^{\chi^0\chi^0}$ and $C^{\tilde g \chi^0}_1$, as a function of $m_{\tilde q_2}$ setting 
$m_{\tilde g}=1.5$ TeV, $M_2=0.4$ TeV and $m_{\tilde q_1}=0.8$ TeV.}
\label{fig:WCs}
\end{center}
\end{figure*}

In Fig.~\ref{fig:WCs}, we show the size of the various contributions to $C_1$ that are 
$C^{\g\g}_1$, $C^{\chi^+\chi^+}_1$, $C_1^{\chi^0\chi^0}$ and $C^{\tilde g \chi^0}_1$
as a function of the squark mass $m_{\tilde q_2}$. For definiteness we set the other
parameters as $m_{\tilde g}=1.5$ TeV, $M_2=0.4$ TeV and $m_{\tilde q_1}=0.8$ TeV.
As already anticipated, $C^{\chi^+\chi^+}_1$ and $C_1^{g\chi^0}$ dominate over 
$C^{\g\g}_1$ in large regions of the parameter space. 
By contrast, the pure neutralino effects encoded in $C_1^{\chi^0\chi^0}$ are always 
very suppressed and therefore negligible. 

\begin{figure*}
%\begin{figure*}[t!]
\begin{center}
%\begin{tabular}{cc}
\includegraphics[scale=1.0]{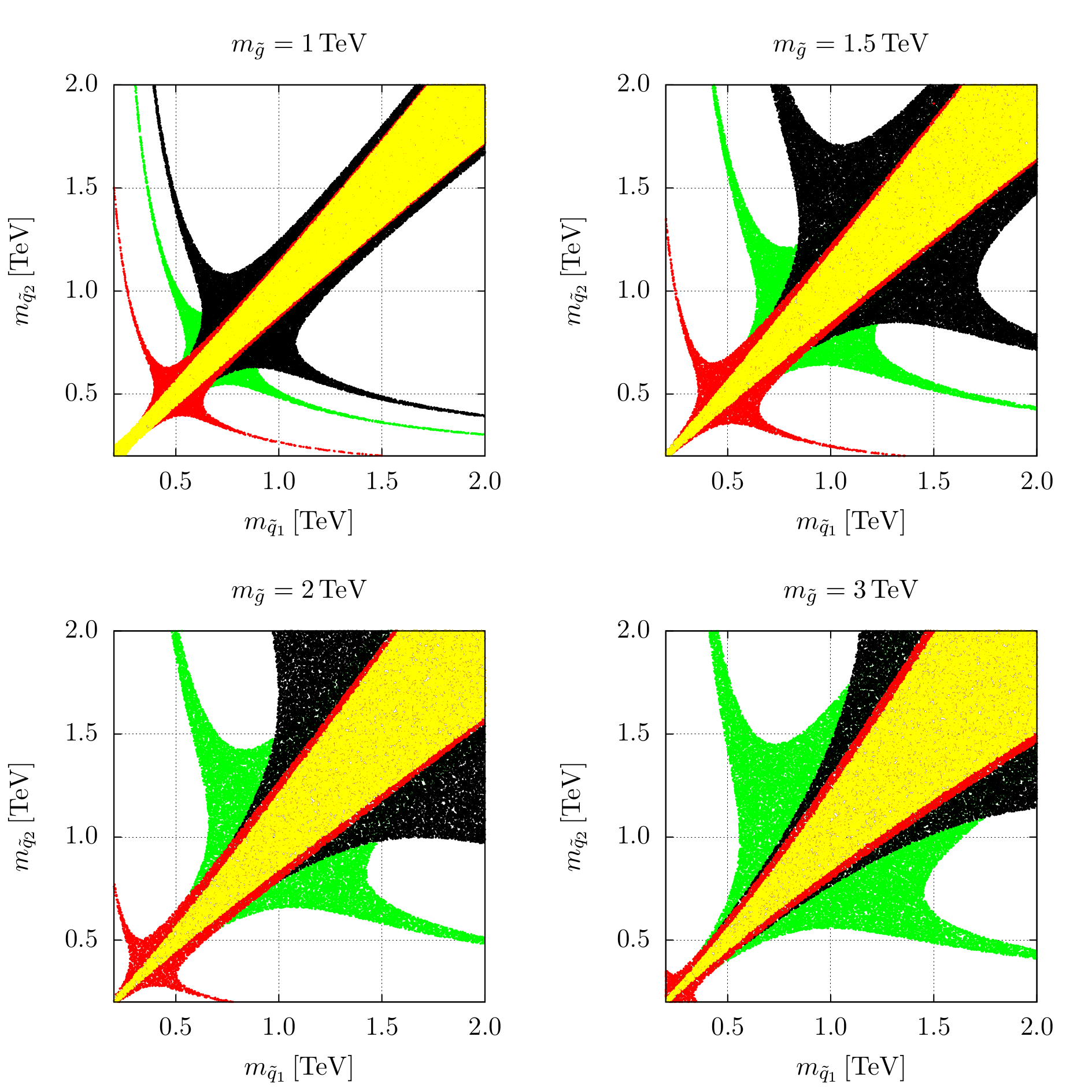}
%\end{tabular}
\caption{Allowed regions in the squark mass plane
for $m_\g = 1~$TeV (upper left), $m_\g = 1.5~$TeV (upper right),
$m_\g = 2~$TeV (lower left), $m_\g = 3~$TeV (lower right).
Different colours correspond to: $M_2 = 100~$GeV (yellow),
$M_2 = 200~$GeV (red), $M_2 = 400~$GeV (green), $M_2 = 1000~$GeV (black).}
\label{fig:exclusion}
\end{center}
\end{figure*}

In Fig.~\ref{fig:exclusion}, we show
the allowed regions in the squark mass plane for $m_\g=1~$TeV (upper left),
$m_\g=1.5~$TeV (upper right), $m_\g = 2~$TeV (lower left), $m_\g=3~$TeV (lower right).
The various colours correspond to: $M_2 = 100~$GeV (yellow), $M_2=200~$GeV (red),
$M_2 = 400~$GeV (green), $M_2 = 1000~$GeV (black). Here, we have neglected the 
mixings in the chargino and neutralino mass matrices keeping only the dominant pure 
Wino contribution (see the following section for more details).  On general grounds, from
Fig.~\ref{fig:exclusion} we learn that there is a very significant sensitivity on
the Wino mass $M_2$. In turn, this means that chargino/neutralino effects are
extremely important and therefore their inclusion in phenomenological analyses of
SUSY alignment models is mandatory.

%%%%%%%%%%%%%%%%%%%%%%%%%%%%%%%%%%%%%%%%%%%%%%%%%%%%%%%%%%%%%%%%%%
%%%%%%%%%%%%%%%%%%%%%%%%%%%%%%%%%%%%%%%%%%%%%%%%%%%%%%%%%%%%%%%%%%

%%%%%%%%%%%%%%%%%%%%%%%%%%%%%%%%%%%%%%%%%%%%%%%%%%%%%%%%%%%%%%%%%%
\subsection{Approximated results}
\label{sec:approximated_results}

%%%%%%%%%%%%%%%%%%%%%%%%%%%%%%%%%%%%%%%%%%%%%%%%%%%%%%%%%%%%%%%%%%
In the following, we provide the expressions for $C^{\g\g}_1$, $C^{\chi^+\chi^+}_1$,
$C_1^{\chi^0\chi^0}$, $C^{\g \chi^0}_1$ in many useful limits.

%%%%%%%%%%%%%%%%%%%%%%%%%%%%%%%%%%%%%%%%%%%%%%%%%%%%%%%%%%%%%%%%%%
\subsubsection{No chargino/neutralino mixing}
\label{sec:no_mixing}
%%%%%%%%%%%%%%%%%%%%%%%%%%%%%%%%%%%%%%%%%%%%%%%%%%%%%%%%%%%%%%%%%%

The expressions of the Wilson coefficients $C^{\chi^+\chi^+}_1$, $C_1^{\chi^0\chi^0}$ and $C^{\g \chi^0}_1$ of 
Eqs.~(37)-(39), depend on the chargino and neutralino diagonalization matrices $Z_{-}$ and $Z_{N}$. 
In the unbroken $SU(2)$ limit, where there is no gaugino mixing, we are left with the pure exchange of Higgsinos, 
Wino and Bino. However, Higgsino and Bino effects are both extremely suppressed by light Yukawas and small 
$U(1)_Y$ gauge coupling, respectively. After $SU(2)$ breaking, Higgino/Wino mixings will induce corrections 
to the pure Wino contribution of order $v^2/{\rm max}(\mu^2,M^2_2)$ which are sizable only for relatively light
Higgsinos and Winos.
Thus, the leading chargino/neutralino and gluino-neutralino contributions, as obtained
by neglecting chargino/neutralino mixings and $U(1)_Y$ interactions, are given by the
compact expressions
\begin{align}
\label{eq:no_chimixing1}
C^{\chi^+\chi^+}_1 &= - \frac{\alpha^2_w}{2} \left(s_{\mysmall L}c_{\mysmall L}e^{i\phi_{\mysmall L}}\right)^2
C_{\tilde{d}_L\tilde{d}_L}(M^2_2, M^2_2)~, 
\\
C_1^{\chi^0\chi^0} &= - \frac{\alpha^2_w}{8} \left(s_{\mysmall L}c_{\mysmall L}e^{i\phi_{\mysmall L}}\right)^2 
\bigg( \frac{1}{2} M_2^2~B_{\tilde{u}_L\tilde{u}_L}(M^2_2, M^2_2) + C_{\tilde{u}_L\tilde{u}_L}(M^2_2, M^2_2)\bigg)\, ,
\\
\label{eq:no_chimixing2}
C^{\tilde g \chi^0}_1 &= -\frac{\alpha_s \alpha_w}{3}
\left(s_{\mysmall L}c_{\mysmall L}e^{i\phi_{\mysmall L}}\right)^2 
\bigg( \frac{1}{2} M_2 m_\g~B_{\tilde{u}_L\tilde{u}_L}(M^2_2,m^2_\g) + C_{\tilde{u}_L\tilde{u}_L}(M^2_2,m^2_\g) \bigg)\,.
\end{align}
Eqs.~(\ref{eq:no_chimixing1})-(\ref{eq:no_chimixing2}) together with the expressions of $C^{\g\g}_{1,4,5}$ and 
${\tilde C}^{\g\g}_{1}$ of sec.~\ref{sec:full_results}, provide the full set of Wilson coefficients describing $D^0-\bar D^0$ mixing. 
These expressions, which provide an excellent approximation of the full results of sec.~\ref{sec:full_results},
are entirely expressed in terms of physical parameters, i.e. masses, mixing angles and CPV phases, and do not require 
any numerical diagonalization of the squark and chargino/neutralino mass matrices to be used.
%
%
%%%%%%%%%%%%%%%%%%%%%%%%%%%%%%%%%%%%%%%%%%%%%%%%%%%%%%%%%%%%%%%%%%
%\clearpage
\subsubsection{Quasi-degenerate squarks}
\label{sec:degenerate}
%%%%%%%%%%%%%%%%%%%%%%%%%%%%%%%%%%%%%%%%%%%%%%%%%%%%%%%%%%%%%%%%%%
%
In the following, we provide the approximate expressions for the Wilson coefficients of sec.~\ref{sec:full_results} 
in the limit of quasi-degenerate squarks $m_{\tilde{q}_1}\simeq m_{\tilde{q}_2} \equiv m_{\tilde{q}}$ 
($\tilde{q}_1$ and $\tilde{q}_2$ are the two squarks running in the loop) neglecting the chargino 
and neutralino mixings. We find that
\begin{align}
\left(C^{\g\g}_1\right)_{\rm deg} &=
-\frac{\alpha^2_s}{216 m_{\tilde{q}}^{2}} \left(\delta_{u}^{L}\right)^2_{21}\bigg( 24 x_{gq} f_6(x_{gq}) + 66 {\tilde f}_6(x_{gq}) \bigg)\, ,
\\
(C^{\chi^+\chi^+}_1)_{\rm deg} &= -\frac{\alpha^2_w}{8 m_{\tilde{q}}^{2}}{\tilde f}_6(x_{wq}) \left(\delta_{u}^{L}\right)^2_{21} \, ,
\\
(C_1^{\chi^0\chi^0})_{\rm deg} &=
-\frac{\alpha^2_w}{16 m_{\tilde{q}}^{2}} \left(\delta_{u}^{L}\right)^2_{21} \bigg( x_{wq} f_6(x_{wq}) + \frac{1}{2}{\tilde f}_6(x_{wq}) \bigg) \, , 
\\
( C^{\g \chi^0}_1 )_{\rm deg} &=
-\frac{\alpha_s \alpha_w}{6m_{\tilde{q}}^{2}} \left(\delta_{u}^{L}\right)^2_{21} 
\bigg(\sqrt{x_w x_g}~f_6(x_w, x_g) + \frac{1}{2}{\tilde f}_6(x_w, x_g)\bigg) \, ,  
\\
({\tilde C}^{\g\g}_1)_{\rm deg} &=
-\frac{\alpha^2_s}{216 m_{\tilde{q}}^{2}} \left(\delta_{u}^{R}\right)^2_{21} 
\bigg( 24 x_{gq} f_6(x_{gq}) + 66 {\tilde f}_6(x_{gq}) \bigg) \, , 
\\
\left(C^{\g\g}_4\right)_{\rm deg} &= -\frac{\alpha^2_s}{3 m_{\tilde{q}}^{2}} 
\left(\delta_{u}^{L}\right)_{21}\left(\delta_{u}^{R}\right)_{21} \bigg( 7 x_{gq} f_6(x_{gq}) - {\tilde f}_6(x_{gq}) \bigg) \, , 
\\
\left(C^{\g\g}_5\right)_{\rm deg} &= -\frac{\alpha^2_s}{9 m_{\tilde{q}}^{2}} 
\left(\delta_{u}^{L}\right)_{21}\left(\delta_{u}^{R}\right)_{21} \bigg( x_{gq} f_6(x_{gq}) +5 {\tilde f}_6(x_{gq}) \bigg) \, , 
\label{eq:quasi_degenerate}
\end{align}
where $x_{gq} = m_\g^2/m_{\tilde{q}}^{2}$, $x_{wq} = M_2^2/m_{\tilde{q}}^{2}$, and the loop functions $f_6(x)$,
${\tilde f}_6(x)$, $f_6(x,y)$, and ${\tilde f}_6(x,y)$ are given in the appendix. The above expressions extend  
the results of Gabbiani et al.~\cite{Gabbiani:1996hi} where only the pure gluino contributions were considered.
%
%
%%%%%%%%%%%%%%%%%%%%%%%%%%%%%%%%%%%%%%%%%%%%%%%%%%%%%%%%%%%%%%%%%%%%
\subsubsection{Split squarks}
\label{sec:split}
%%%%%%%%%%%%%%%%%%%%%%%%%%%%%%%%%%%%%%%%%%%%%%%%%%%%%%%%%%%%%%%%%%%%

When the squark mass splittings are sizable, the results obtained in the MI approximation are not trustable. 
As an illustrative example, we consider the limit of split squark families where it is assumed that the heaviest 
squark is completely decoupled, i.e. $m_{\tilde{q}_1} \to \infty$.
In this scenario, the prescriptions for $(\delta_{u}^{L})_{21}$ and $(\delta_{u}^{R})_{21}$ are
\begin{align}
(\delta_{u}^{L})_{21} &= \frac{(M^{2}_{{\tilde u}, LL})_{21}}{m_{\tilde{q}_1}^{2}}
\simeq \frac{(M^{2}_{{\tilde d}, LL})_{21}}{m_{\tilde{q}_1}^{2}} - \lambda_{\rm C}
\equiv (\delta_{d}^{L})_{21} - \lambda_{\rm C} \, , 
\\
(\delta_{u}^{R})_{21} &\simeq
\frac{(M^{2}_{{\tilde u}, RR})_{21}}{m_{\tilde{u}_1}^{2}}~.
\label{defdel_split}
\end{align}
Starting again from the full results of sec.~\ref{sec:full_results}, we end up with the following expressions
\begin{align}
(C^{\g\g}_1)_{\rm split} &= -\frac{\alpha^2_s}{9 m_{\tilde{q}_2}^{2}}
\left(\delta_{u}^{L}\right)^2_{21} \bigg( x_{gq}~D_0(x_{gq}) + 11 D_2(x_{gq}) \bigg)\, ,
\\
%\end{align}
%
%\begin{align}
(C^{\chi^+\chi^+}_1)_{\rm split} &= - \frac{\alpha^2_w}{2m_{\tilde{q}_2}^{2}}
D_2(x_{wq})\left(\delta_{u}^{L}\right)^2_{21}~,
\\
(C_1^{\chi^0\chi^0})_{\rm split} &= - \frac{\alpha^2_w}{8m_{\tilde{q}_2}^{2}}
\left(\delta_{u}^{L}\right)^2_{21} \bigg( \frac{x_{wq}}{2}~D_0(x_{wq}) + D_2(x_{wq}) \bigg)~,
\\
(C^{\g \chi^0}_1)_{\rm split} &= -\frac{\alpha_s \alpha_w}{3m_{\tilde{q}_2}^{2}}
\left(\delta_{u}^{L}\right)^2_{21} \bigg(\frac{\sqrt{x_{wq}x_{gq}}}{2}~D_0(x_{wq},x_{gq}) + D_2(x_{wq},x_{gq})\bigg)~,
\\
%\end{align}
%
%\begin{align}
({\tilde C}^{\g\g}_1)_{\rm split} &= -\frac{\alpha^2_s}{9 m_{\tilde{q}_2}^{2}}
\left(\delta_{u}^{R}\right)^2_{21}  \bigg( x_{gq}~D_0(x_{gq}) + 11 D_2(x_{gq}) \bigg)\, ,
\\
(C^{\g\g}_4)_{\rm split} &= -\frac{\alpha^2_s}{3 m_{\tilde{q}_2}^{2}}
\left(\delta_{u}^{L}\right)_{21}\left(\delta_{u}^{R}\right)_{21} \bigg( 7x_{gq}~D_0(x_{gq}) - 4 D_2(x_{gq}) \bigg)~,
\\
(C^{\g\g}_5)_{\rm split} &= -\frac{\alpha^2_s}{9 m_{\tilde{q}_2}^{2}}
\left(\delta_{u}^{L}\right)_{21}\left(\delta_{u}^{R}\right)_{21} \bigg( x_{gq}~D_0(x_{gq}) + 20 D_2(x_{gq}) \bigg)~,
\label{eq:split}
\end{align}
where $x_{gq}=m_\g^2/m_{\tilde{q}_{2}}^{2}$, $x_{wq}=M_2^2/m_{\tilde{q}_{2}}^{2}$,
and the loop functions $D_{0,2}(x)$ are defined in the appendix.

In the limit of $m_{\tilde{q}_1} \to \infty$, the Wilson coefficients of Eqs.~(46)-(52) vanish since they decouple with 
$m_{\tilde q} = (m_{\tilde q_1} + m_{\tilde q_2})/2$ while those of Eqs.~(55)-(61) do not. This clearly shows that the 
expressions of Eqs.~(46)-(52) are completely inadequate to describe scenarios with large squark mass splittings, 
as expected.

%%%%%%%%%%%%%%%%%%%%%%%%%%%%%%%%%%%%%%%%%%%%%%%%%%%%%%%%%%%%%%%%%%
\begin{figure*}[t!]
\includegraphics[width=0.5\textwidth]{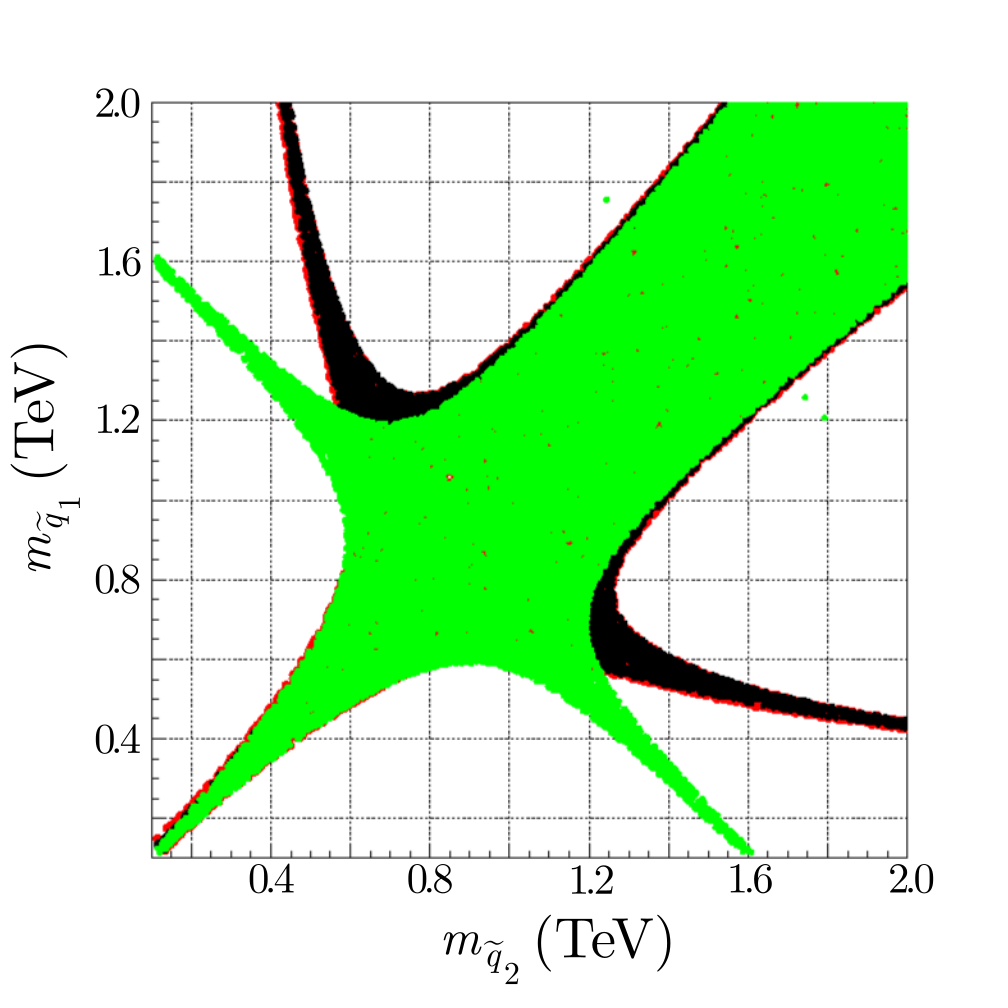}~~~
\includegraphics[width=0.5\textwidth]{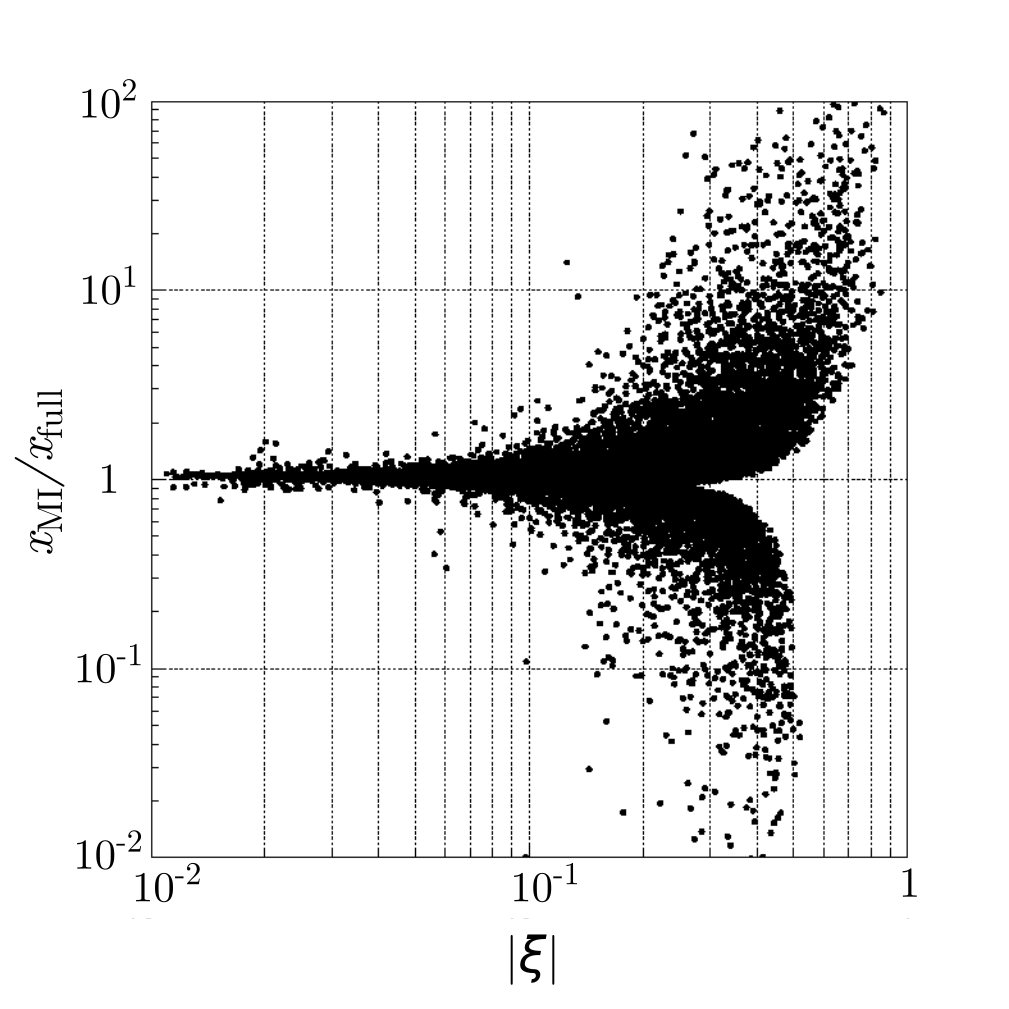}
\caption{
Left: allowed regions in the squark mass plane using the full
computation (red points), neglecting neutralino/chargino mixings 
(black points) and in the MI approximation (green points).
Right: full over MI approximation results for $x$ vs. $|\xi|$.
}
\label{fig:anatomy}
\end{figure*}
%%%%%%%%%%%%%%%%%%%%%%%%%%%%%%%%%%%%%%%%%%%%%%%%%%%%%%%%%%%%%%%

In Fig.~\ref{fig:anatomy}, we compare our full results as obtained working in the squark mass basis with the MI results in
the case of quasi-degenerate squarks. The plots of Fig.~\ref{fig:anatomy} have been obtained for $m_{\tilde g} = 1.5~$TeV, 
$0.2~{\rm TeV}\leq (M_2,\mu)\leq 1~{\rm TeV}$ and assuming $(\delta_{d}^{L})_{21}=(\delta_{u}^{R})_{21}=0$. 
In the left plot, we show the allowed regions in the squark mass plane using 
the full computation of sec.~\ref{sec:full_results} (red points), neglecting the
neutralino/chargino mixings (black points), see Eqs.~(43)-(45), and in the MI 
approximation (green points), see Eqs.~(46)-(52).

Our numerical results confirm that neutralino/chargino mixing effects are indeed
rather small. Yet, we find that for light Wino and Higgsino, $(M_2,\mu)\lesssim v$,
corrections up to $50\%$ are still possible.
On the other hand, for large squark mass splittings, we observe significant
departures of the MI approximation from the exact results. This is also evident in the right
plot where we show the ratio between $x$ in the MI approximation, $x_{\rm MI}$, and in the full
computation, $x_{\rm full}$, as a function of $|\xi|$: for $|\xi|\lesssim 0.1$ the two computations 
nicely agree while they can differ very significantly for $|\xi| \sim {\mathcal O}(1)$.

Concerning the case of split-squarks, we have explicitly checked that the MI approximation 
formulae reproduce quite accurately the full results provided $|\xi| \gtrsim 0.6$.

%%%%%%%%%%%%%%%%%%%%%%%%%%%%%%%%%%%%%%%%%%%%%%%%%%%%%%%%%%%%%%
%\clearpage
\section{CPV in $D^0-\bar D^0$ mixing}
\label{sec:CPV_mixing}
%%%%%%%%%%%%%%%%%%%%%%%%%%%%%%%%%%%%%%%%%%%%%%%%%%%%%%%%%%%%%%

We are ready now to analyse possible CPV effects for $D^0-\bar D^0$ mixing in SUSY alignment models. 
On general ground, we notice that in the limit of complete alignment, that is for $(\delta_{d}^{L})_{21}=(\delta_{u}^{R})_{21}=0$, 
CPV effects in $D^0-\bar D^0$ mixing are vanishing as $(\delta_{u}^{L})_{21}$, which is the only 
source of flavor violation can be taken to be real without loss of generality~\cite{Altmannshofer:2010ad}.

On the other hand, possible CPV sources stem from $(\delta_{d}^{L})_{21}$ and/or 
$(\delta_{u}^{R})_{21}$. In the former case, CPV effects in $D^0-\bar D^0$ and
$K^0-\bar K^0$ mixings are correlated due to the underlying $SU(2)_L$ symmetry
and the leading effects are generated through the SM operator $Q_1$,
see Eq.~\eqref{eq:DF2_operators}.
By contrast, in the latter case, the effects in $D^0-\bar D^0$ and $K^0-\bar K^0$
mixings are not correlated and the leading effects for $D^0-\bar D^0$ arise typically
from the operator $Q_4$.

For a qualitative understanding of CPV effects in $D^0-\bar D^0$ mixing, it is convenient 
to consider the CPV phase in the mixing in the approximation that the SM contributions are 
dominated by the first two generations and are made real, as explained above,
see Eqs.~\eqref{eq:agamma_asl},\eqref{IM} and~\eqref{AgamNP}. In that case we can bound 
the amount of CPV by setting the contributions of the SM to $M_{12}$ to zero (not allowing for 
accidental cancellations) hence in the following $M_{12}$ is assumed to be totally dominated by 
the NP contributions. In this case we can write (omitting for simplicity the ${\rm SM\in Real}$ 
subscript in (\ref{AgamNP}))
\bea
\frac{A_\Gamma}{x} \simeq \frac{{\rm Im} M_{12}}{{\rm Re} M_{12}}\,.
\label{eq:phi_M}
\eea
Again, we emphasise that it is assumed that $M_{12} = M_{12}^{\rm NP}$, to maximise the contributions.
We are going now to analyse two distinct cases where either $(\delta_{d}^{L})_{21} \neq 0$ and $(\delta_{u}^{R})_{21} = 0$
or $(\delta_{d}^{L})_{21}=0$ and $(\delta_{u}^{R})_{21}\neq 0$ outlying their peculiar phenomenological features.

\begin{enumerate}
\item  $(\delta_{d}^{L})_{21}\neq 0$, $(\delta_{u}^{R})_{21}=0$.
In this case we find that
\bea
\frac{A_\Gamma}{x}
\approx 
\frac{{\rm Im} [(M^{2}_{{\tilde u}, LL})_{21}]}{{\rm Re} [(M^{2}_{{\tilde u}, LL})_{21}]} 
\equiv 
\frac{{\rm Im}\left[ (\delta_{d}^{L})^2_{21} + 8\lambda_{\rm C} \xi (\delta_{d}^{L})_{21}\right]}
{{\rm Re}\left[\left( (\delta_{d}^{L})_{21} + 4\lambda_{\rm C} \xi \right)^2\right]}
\approx
\frac{{\rm Im}(\delta_{d}^{L})_{21}}{2\lambda_{\rm C} \xi}~,
\label{approx_cpv_c1}
\eea
where the first approximation is valid at the leading-order expansion in $(M^{2}_{{\tilde u}, LL})_{21}$ while the last 
one, obtained by using Eqs.~\eqref{qsa_approx}--\eqref{defdel}, is valid for ${\rm Re}(\delta_{d}^{L})_{21}\ll 4\lambda_{\rm C}\xi$.
Interestingly, Eq.~(\ref{approx_cpv_c1}) shows that, for a given value of $x$, the largest effects in $A_\Gamma$
are expected for small values of $\xi$, i.e. for relatively degenerate squarks. 
The maximum value for $A_\Gamma$ is found by imposing the constraints from $\epsilon_K$ and $x$ which have 
the following parametric expressions 
\be
\Delta\epsilon_K  \sim  {\rm Im} \left[(\delta_{d}^{L})^{2}_{21}\right] \, , \qquad
%\\ 
x  \sim  {\rm Re}\bigg[\bigg( (\delta_{d}^{L})_{21} + 4\lambda_{\rm C}\xi\bigg)^{2}\bigg]\,.
\label{constraints}
\ee
In particular, in the quasi-degenerate scenario (see sec.~\ref{sec:degenerate}) and
assuming that gluino effects are dominant, we end up with the following estimates
\be
\label{epsilonK}
\frac{|\Delta\epsilon_K|}{\epsilon_K^{\rm\mysmall SM}} \approx 0.4 \,
%\frac{|{\rm Im} (\delta_{d}^{L})^2_{21}|}{10^{-4}}
\frac{|(\delta_{d}^{L})^2_{21}|}{10^{-4}} \sin(2\phi_L)\!
\left(\frac{1.5\, {\rm TeV}}{{\tilde m}_Q}\right)^{\!2}\!\!, ~~~~~~
x \approx  8 \!\times\! 10^{-3}\!
\left(\!\frac{\xi}{0.2}\!\right)^{\!\!2}\!\!
\left(\frac{1.5 \, {\rm TeV}}{\tilde m_Q}\right)^{\!2}\!\!,
\ee
where we have set $m_g={\tilde m}_Q=1~$TeV, $(\delta_{d}^{L})_{21}=e^{i\phi_L}|(\delta_{d}^{L})_{21}|$, 
and assumed again that ${\rm Re}(\delta_{d}^{L})_{21} \ll 4\lambda_{\rm C} \xi$. Therefore, imposing the 
constraint $|\Delta\epsilon_K|/\epsilon_K^{\rm\mysmall SM} \lesssim 0.4$, and setting $\phi_L=45^\circ$, 
we find the upper bound
\bea
|A_\Gamma| \lesssim 7 \times 10^{-4} \left(\frac{x^{\rm max}}{6.7 \times 10^{-3}}\right)\,,
\eea
as confirmed by the lower plot on the right of Fig.~5.
Notice that $A_\Gamma \sim \sin\phi_L$ while $\Delta\epsilon_K \sim \sin2\phi_L$ and therefore the constraint 
from $\Delta\epsilon_K$ can be relaxed for $\phi_L \approx 90^\circ$ while maximizing $A_\Gamma$.
\item $(\delta_{d}^{L})_{21}=0$ and $(\delta_{u}^{R})_{21}\neq 0$.
In this case we find that
\be
\frac{A_\Gamma}{x}
\approx 400~
\frac{{\rm Im}\left[(\delta_{u}^{L})_{21}(\delta_{u}^{R})_{21}\right]}
{{\rm Re}\left[(\delta_{u}^{L})^{2}_{21}\right]}
 \approx  200~
\frac{{\rm Im}(\delta_{u}^{R})_{21}}{2\lambda_{\rm C} \xi}~,
\label{approx_cpv_c4}
\ee
where the first approximation holds in the limit where
${\rm Re}\left[(\delta_{u}^{L})_{21}(\delta_{u}^{R})_{21}\right] \ll {\rm Re}\left[(\delta_{u}^{L})^{2}_{21}\right]$.
Comparing Eq.~\eqref{approx_cpv_c1} with Eq.~\eqref{approx_cpv_c4}, we learn that $(\delta_{u}^{R})_{21}$ is 
potentially much more effective than $(\delta_{d}^{L})_{21}$ to generate large CPV effects in $D^0-\bar D^0$ mixing.
In particular, for ${\rm Im}(\delta_{u}^{R})_{21}\approx {\rm Im}(\delta_{d}^{L})_{21}$
(notice that in alignment models $(\delta_{u}^{R})_{21}$ might be even larger
than $(\delta_{d}^{L})_{21}$, see Tab.~\ref{QSA}) the effect driven by
$(\delta_{u}^{R})_{21}$ is typically more than two orders of magnitude larger
than that from $(\delta_{d}^{L})_{21}$. The reason of this can be traced back
remembering that $C_4$ is highly enhanced with respect to $C_1$ by a larger
hadronic matrix element, larger QCD-induced RGE effects, and also by a larger
loop function. Moreover, from a pure phenomenological perspective, we remember
that $(\delta_{u}^{R})_{21}$ does not suffer from the $K^0-\bar K^0$ mixing
constraints, in contrast with $(\delta_{d}^{L})_{21}$.

\end{enumerate}

%%%%%%%%%%%%%%%%%%%%%%%%%%%%%%%%%%%%%%%%%%%%%%%%%%%%%%%%%%%%%%%%%%
\begin{figure*}[t]
\includegraphics[width=0.5\textwidth]{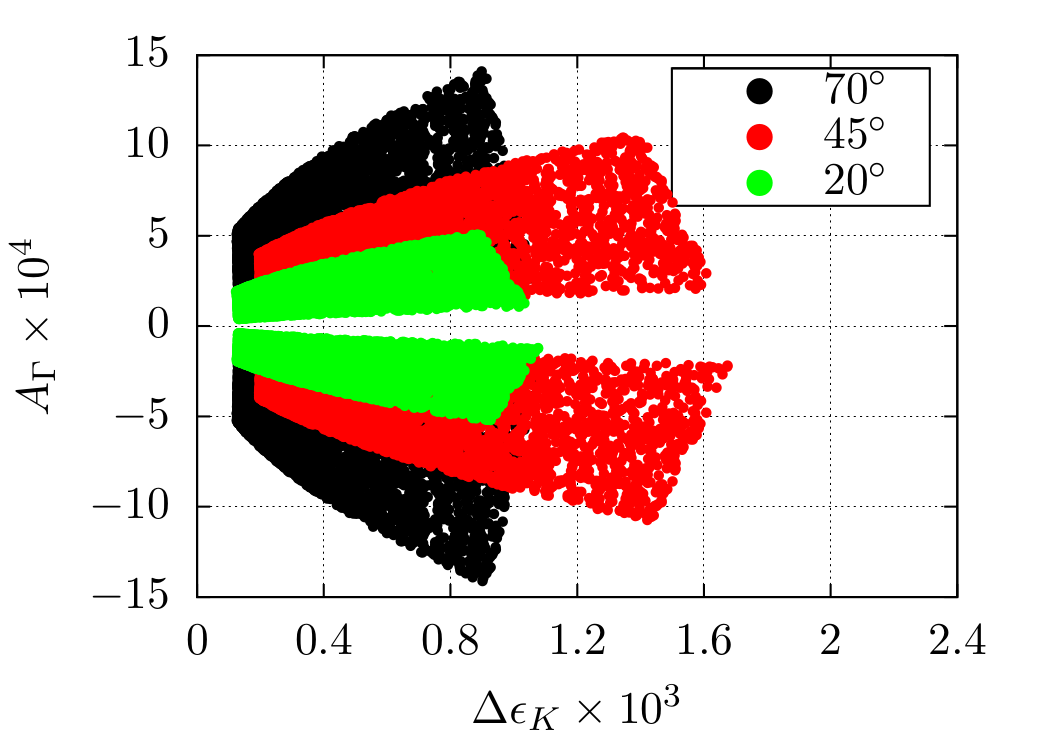}~~~
\includegraphics[width=0.5\textwidth]{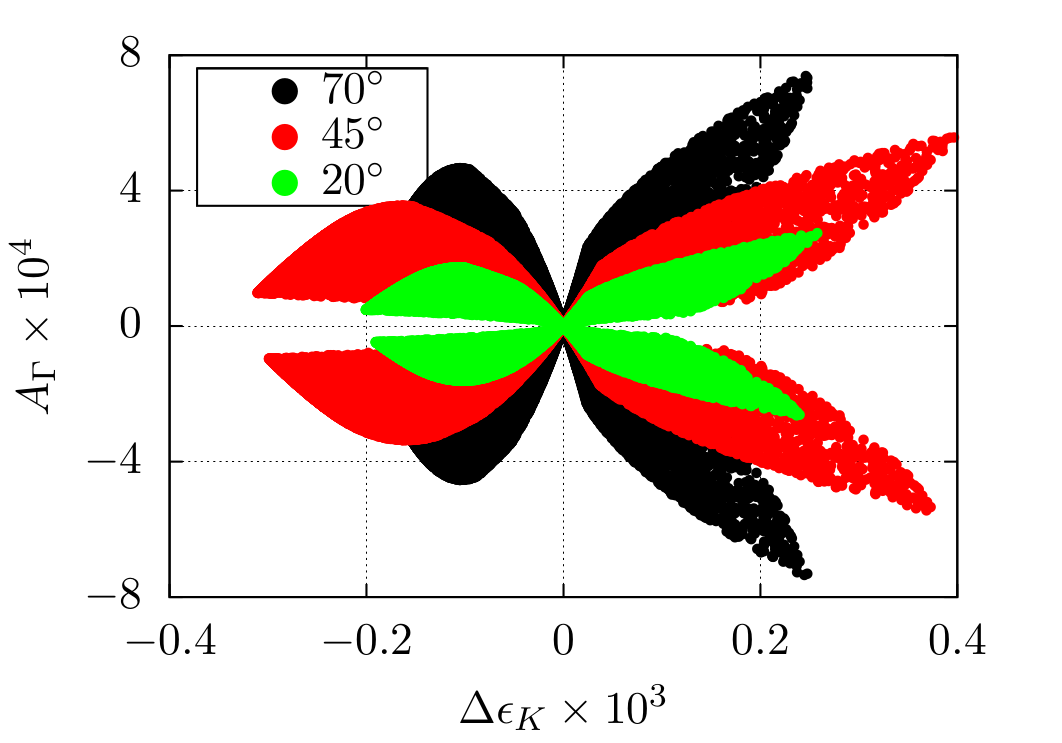}
\includegraphics[width=0.5\textwidth]{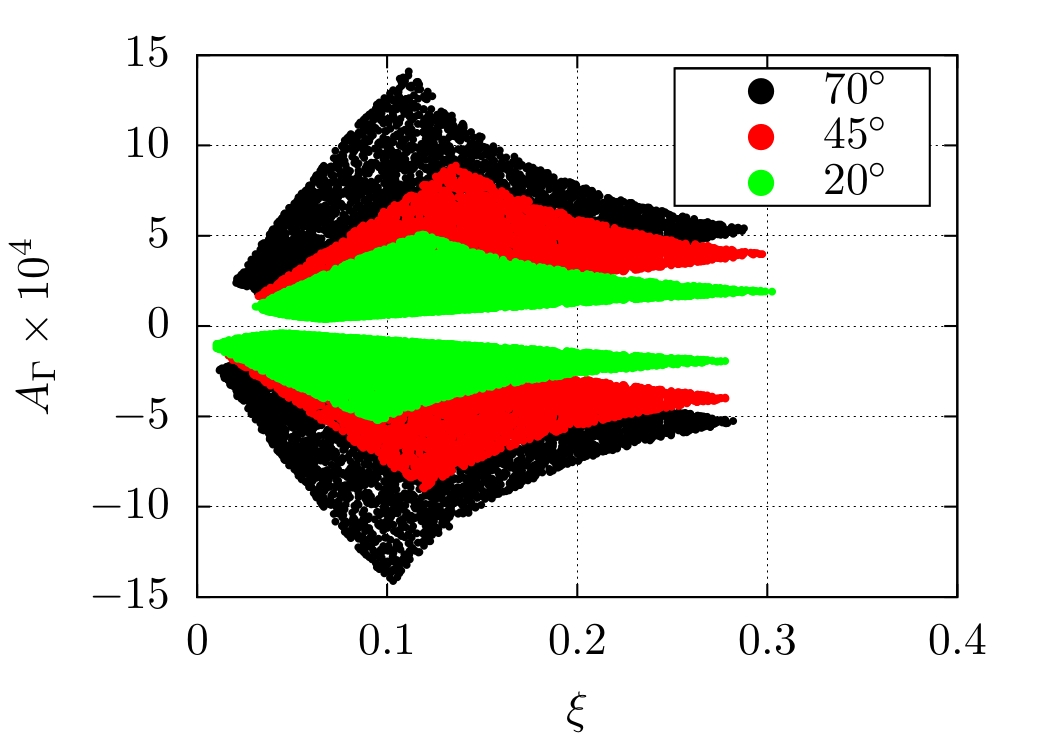}~~~
\includegraphics[width=0.5\textwidth]{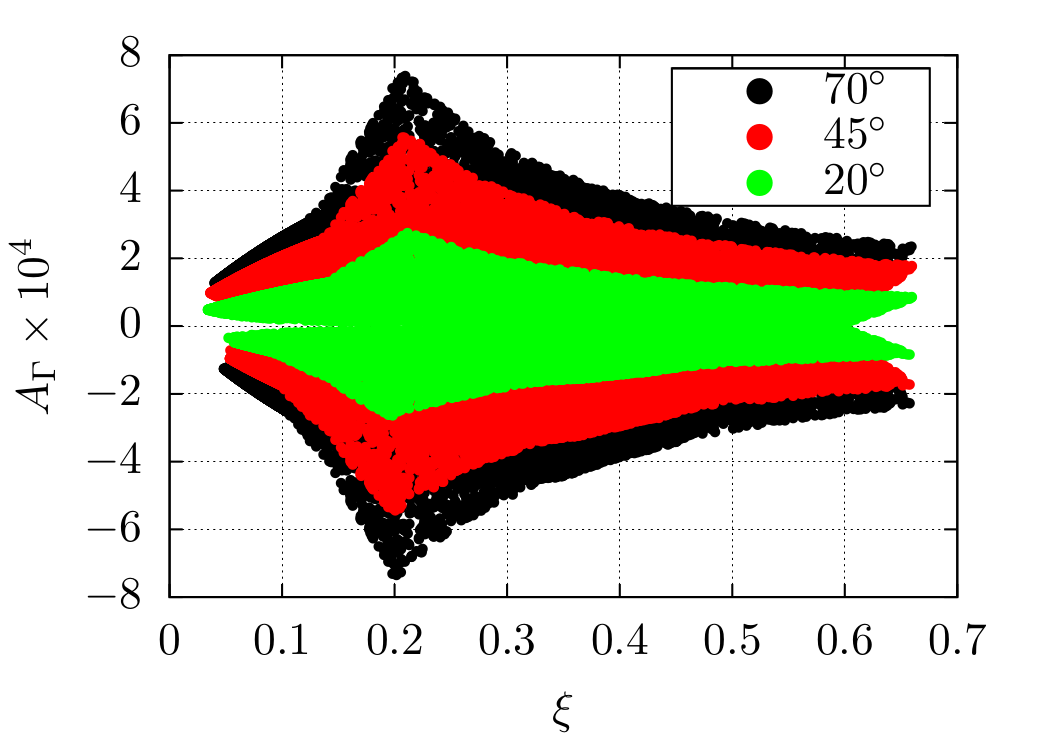}
\caption{
In the upper plots we show $A_\Gamma$ vs. $\Delta\epsilon_K$ while in the lower plots 
$A_\Gamma$ (left) vs. $\xi$. The plots on the left (right) include only EW-ino (gluino) effects. 
Green, red and black points correspond to ${\rm arg} (\delta^{L}_{d})_{21} = 20^{°}, 45^{°}, 70^{°}$, 
respectively.}
\label{plot:cpv}
\end{figure*}

%%%%%%%%%%%%%%%%%%%%          Numerical analysis          %%%%%%%%%%%%%%%%%%%%

In Fig.~\ref{plot:cpv}, we show the predictions for $A_\Gamma$ vs. $\Delta\epsilon_K$ (upper plots)
and $A_\Gamma$ vs. $\xi$ (lower plots) in the case 1. The plots on the left (right) include only EW-ino (gluino) 
effects. Green, red and black points correspond to ${\rm arg} (\delta^{L}_{d})_{21} = 20^{°}, 45^{°}, 70^{°}$, 
respectively. An intriguing feature emerging by these plots is the growth of $A_\Gamma$ for decreasing values 
of $\xi$ which might be traced back from Eq.~\eqref{approx_cpv_c1}. Given the collider bounds on 
$m_{\tilde{q}_1} \gtrsim 1$TeV, this implies that $A_\Gamma$ is maximum for $m_{\tilde{q}_2}\approx 1$TeV, 
well above the current experimental bound from direct search.
Moreover, the maximum values for $A_\Gamma$ are reached for ${\rm arg} (\delta^{L}_{d})_{21}$ approaching
$90^\circ$ as in this case the indirect constraint from $\Delta\epsilon_K$ can be relaxed, as already discussed. 
%

%%%%%%%%%%%%%%%%%%%%%%%%%%%%%%%%%%%%%%%%%%%%%%%%%%%%%%%%%
\section{Conclusions}
\label{sec:conclusions}
%%%%%%%%%%%%%%%%%%%%%%%%%%%%%%%%%%%%%%%%%%%%%%%%%%%%%%%%%

In spite of the remarkable success of the SM in describing all the available flavor data on $K$ and $B_{d,s}$ systems, 
it is still possible that New Physics (NP) affects the up-quark sector in a significant manner. This is the case for instance 
of models of alignment, in which the flavor structure of the NP does not satisfy two-generation universality. 
In this work, we have revisited the phenomenology of alignment models both model-independently and within 
supersymmetric scenarios. Assuming that NP contributes to $K^0-\bar K^0$ and $D^0-\bar D^0$ mixings only 
through non-renormalizable operators involving $SU(2)_L$ quark-doublets,
%Assuming that NP contributions to $K^0-\bar K^0$ and $D^0-\bar D^0$ mixings are approximately $SU(2)_L$ invariant, 
we have derived model-independent 
upper bounds on CP violating effects in $D$ meson system. Interestingly enough, we have found that the 
current experimental resolutions are starting to probe the natural predictions of alignment models. 
Our main finding is that within the above framework the bound from $\epsilon_K$ and the current 
value of $x$ (see Tab.~1) constrain CP violation in the $D-\bar D$ mixing to below the per-mil level, 
$A_{\Gamma}\lesssim 0.1\%$ (see Fig.~1 and Eq.~\eqref{eq:mod_ind_bound_obs2}).

Concerning supersymmetric scenarios, in the following we summarize our main results.
\begin{itemize}
\item[{\bf i)}] We have computed the full set of contributions (including pure gluino, mixed neutralino/gluino, 
chargino, and neutralino contributions) for the $D^0-\bar D^0$ mixing amplitude. We have found that chargino 
effects dominate over the pure gluino contribution, which is often the only effect considered in the literature, 
in large regions of the parameter space (see Figs.~2,\,3). Therefore, their inclusion 
in phenomenological analyses of SUSY alignment models is mandatory.
\item[{\bf ii)}] Assuming complete alignment, the second squark generation might be relatively light at the 
level of $m_{\tilde{q}_2}\gtrsim 400$GeV, even for $m_{\tilde{q}_1}\gtrsim 1$TeV (see Fig.~\ref{fig:exclusion}).
\item[{\bf iii)}] CP violating effects in the mixing, described by the quantity $A_\Gamma$ (see Eq.~\eqref{eq:DYf}) and by the semileptonic 
asymmetry $a_{SL}$, which is correlated model-independently with $A_\Gamma$ (see Eq.~(9)) can saturate the current experimental bound 
while being compatible with all flavor and collider constraints, see Fig.~\ref{plot:cpv}. 
Interestingly, the largest CPV effects are expected for relatively degenerate squarks and therefore for $m_{\tilde{q}_2}\approx m_{\tilde{q}_1}\gtrsim 1$TeV. 
In particular, assuming that NP contributions to $K^0-\bar K^0$ and $D^0-\bar D^0$ mixings are approximately $SU(2)_L$ invariant, 
CP violation in D meson systems can saturate our model-independent upper bound (see Eq.~\eqref{eq:mod_ind_bound_obs2}).
\item  [{\bf iv)}] Finally, we have clarified the limit of applicability of the commonly used MI approximation comparing the results of the full and 
MI computations in two relevant squark mass regimes: the quasi-degeneracy and split scenarios. In the former case, the MI approximation 
works well up to squark mass splittings such that $|\xi|\lesssim 0.1$ (see Eq.~(\ref{eq:xi})). 
On the other hand, already for $|\xi| \gtrsim 0.1$, significant departures from the exact results occur which might become dramatic 
for $|\xi| \sim {\mathcal O}(1)$ (see the right plot of Fig.~\ref{fig:anatomy}). In this latter case, the expressions of the split scenario 
reproduce well the full results. For intermediate squark-mass regimes, in particular for $0.1 \lesssim |\xi| \lesssim 0.6$, our full expressions 
of Eqs.~(36), (40)-(45) are highly recommended.

\end{itemize}
%

%\footnotesize

\subsubsection*{Acknowledgments}
The research of PP is supported by the ERC Advanced Grant No.  267985 (DaMeSyFla), by the research grant TAsP 
(Theoretical Astroparticle Physics), and by the Istituto Nazionale di Fisica Nucleare (INFN). PP thanks G.  Buchalla, 
G. Isidori, U. Nierste  and J. Zupan for the invitation to the MIAPP workshop flavor 2015: New Physics at High Energy 
and High Precision,  where part of his work was performed. GP is supported by the  BSF, ISF, and ERC-2013-CoG grant 
(TOPCHARM \# 614794) and acknowledge discussions with A. Kadosh.

%%%%%%%%%%%%%%%%%%%%%%%%%%%%%%%%%%%%%%%%%%%%%%%%%%%%%%%%%%%%%%%%%%%
\section{Appendix A}
%%%%%%%%%%%%%%%%%%%%%%%%%%%%%%%%%%%%%%%%%%%%%%%%%%%%%%%%%%%%%%%%%%%

In the following, we specify the notation used in the text for the squark and chargino/neutralino 
mass matrices. Under the approximations outlined in sec.~\ref{sec:full_results}, we can perform an 
exact diagonalization of the $2 \times 2$ squark mass matrices $M^{2}_{\tilde{q}, LL}$ and 
$M^{2}_{\tilde{u}, RR}$ by means of the unitary matrices $U_L$ and $U_R$ defined as
\begin{eqnarray}
U^{\dagger}_L M^{2}_{\tilde{q}, LL} U_L &=& {\rm diag}(m^2_{\tilde q_1},m^2_{\tilde q_2}) \, , \nn \\
U^{\dagger}_R M^{2}_{\tilde{u}, RR} U_R &=& {\rm diag}(m^2_{\tilde u_1},m^2_{\tilde u_2}) \, ,
\end{eqnarray}
where $U_L$ and $U_R$ read
\begin{eqnarray}
U_L =
%&=& 
\begin{pmatrix} c_{\mysmall L} & -s_{\mysmall L} e^{-i\phi_{\mysmall L}} 
\\
s_{\mysmall L} e^{i\phi_{\mysmall L}} & c_{\mysmall L} \end{pmatrix} \, , 
\qquad\qquad
%\nn    \\
U_R =
%&=& 
\begin{pmatrix} c_{\mysmall R} & -s_{\mysmall R} e^{-i\phi_{\mysmall R}} \\
s_{\mysmall R} e^{i\phi_{\mysmall R}} & c_{\mysmall R} \end{pmatrix} \, .
\end{eqnarray}
The flavor mixing angles $s_{L,R}$, $c_{L,R}$ and the CPV phases $\phi_{L,R}$
are defined as
\begin{eqnarray}
s_{\mysmall L}c_{\mysmall L}e^{i\phi_{\mysmall L}} =
%&=&
\frac{(M^{2}_{\tilde{q}, LL})_{21}}{(m_{\tilde{q}_1}^{2}-m_{\tilde{q}_2}^{2})} \, , \qquad
%\nn \\
s_{\mysmall R}c_{\mysmall R}e^{i\phi_{\mysmall R}} =
%&=&
\frac{(M^{2}_{\tilde{u}, RR})_{21}}{(m_{\tilde{u}_1}^{2}-m_{\tilde{u}_2}^{2})} \, ,
\label{eq:mixing_angles}
\end{eqnarray}
while the squark masses read
\begin{equation}
m^2_{\tilde f_{1,2}}=\frac{m^2_{11}+m^2_{22}\pm\sqrt{(m^2_{11}-m^2_{22})^2+4|m_{12}^2|^2}}{2}\,,
\end{equation}
where $f = q,u$ and $m^2_{ij}$ stands for $(M^{2}_{\tilde{q}, LL})_{ij}$
or $(M^{2}_{\tilde{u}, RR})_{ij}$ when $f = q,u$, respectively.

For the chargino and neutralino mass matrices, we have
\be
\left(
\begin{array}{cc}
M_{\chi_1^\pm} & 0 \\
0 & M_{\chi_2^\pm}
\end{array}
\right) =
Z_-^T
\left(
\begin{array}{cc}
M_2                   & \sqrt{2} s_\beta m_W  \\
\sqrt{2} c_\beta m_W  &         \mu
\end{array}
\right)
Z_+ ~,
\label{MCha}
\ee
and
\begin{eqnarray}
{\rm diag}(M_{\chi_1^0},M_{\chi_2^0},M_{\chi_3^0},M_{\chi_4^0})  =  
Z_N^T \, \mathcal{M}_{\chi^0} \, Z_N ~,
\end{eqnarray}
where
\begin{eqnarray} 
\mathcal{M}_{\chi^0} = 
\left(
\begin{array}{cccc}
M_1               &    0             & -c_\beta s_W m_Z   &   s_\beta s_W m_Z   \\
0                 &   M_2            &  c_\beta c_W m_Z   &  -s_\beta c_W m_Z   \\
-c_\beta s_W m_Z  & c_\beta c_W m_Z  &         0          &       -\mu          \\
 s_\beta s_W m_Z  &-s_\beta c_W m_Z  &       -\mu         &         0
\end{array}
\right)\, . \nn \\
\label{MNeu}
\end{eqnarray}
The unitary matrices $Z_{\pm}$ and $Z_N$ are such that the chargino and neutralino
eigenvaules are positive and ordered as $M_{\chi_1} < M_{\chi_2}$ and
$M_{\chi_1^0} < M_{\chi_2^0}< M_{\chi_3^0} < M_{\chi_4^0}$, respectively.

%%%%%%%%%%%%%%%%%%%%%%%%%%%%%%%%%%%%%%%%%%%%%%%%%%%%%%%%%%%%%%%%%%%%
\section{Appendix B}
%%%%%%%%%%%%%%%%%%%%%%%%%%%%%%%%%%%%%%%%%%%%%%%%%%%%%%%%%%%%%%%%%%%%
In the following, we report the loop functions used in the text for the full computation in the 
mass eigenstates:
\begin{align}
B_{\tilde{f} \tilde{f}}(m_3^2,m_4^2) &=
D_0(m_{\tilde{f}_1}^{2},m_{\tilde{f}_1}^{2},m_3^2,m_4^2)-
D_0(m_{\tilde{f}_1}^{2},m_{\tilde{f}_2}^{2},m_3^2,m_4^2) 
+ \{m_{\tilde{f}_1}^{2} - m_{\tilde{f}_2}^{2} \}\,,
\\
%& + 
%D_0(m_{\tilde{f}_2}^{2},m_{\tilde{f}_2}^{2},m_3^2,m_4^2)-
%D_0(m_{\tilde{f}_2}^{2},m_{\tilde{f}_1}^{2},m_3^2,m_4^2) \\
%
C_{\tilde{f} \tilde{f}}(m_3^2,m_4^2) &=
D_2(m_{\tilde{f}_1}^{2},m_{\tilde{f}_1}^{2},m_3^2,m_4^2)-
D_2(m_{\tilde{f}_1}^{2},m_{\tilde{f}_2}^{2},m_3^2,m_4^2) 
+ \{m_{\tilde{f}_1}^{2} - m_{\tilde{f}_2}^{2} \}\,,
%\\
%&+ 
%D_2(m_{\tilde{f}_2}^{2},m_{\tilde{f}_2}^{2},m_3^2,m_4^2)-
%D_2(m_{\tilde{f}_2}^{2},m_{\tilde{f}_1}^{2},m_3^2,m_4^2)
\end{align}
where $D_{0,2}$ are the standard 4-point functions given by
\begin{align}
D_0(m_1^2,m_2^2,m_3^2,m_4^2) &=  \frac{m_2^2}{(m_2^2-m_4^2) (m_2^2-m_1^2) (m_2^2-m_3^2)} \log \left(\frac{m_1^2}{m_2^2}\right) 
\nn \\
&+\frac{m_3^2}{(m_3^2-m_4^2) (m_3^2-m_1^2) (m_3^2-m_2^2)} \log \left(\frac{m_1^2}{m_3^2}\right) 
\nn \\
&+ \frac{m_4^2}{(m_4^2-m_1^2) (m_4^2-m_2^2) (m_4^2-m_3^2)} \log \left(\frac{m_1^2}{m_4^2}\right)\,,
\end{align}
\begin{align}
D_2(m_1^2,m_2^2,m_3^2,m_4^2) &= \frac{m_2^4}{4(m_2^2-m_4^2) (m_2^2-m_1^2) (m_2^2-m_3^2)} \log \left(\frac{m_1^2}{m_2^2}\right) \nn \\
&+\frac{m_3^4}{4(m_3^2-m_4^2) (m_3^2-m_1^2) (m_3^2-m_2^2)} \log \left(\frac{m_1^2}{m_3^2}\right) \nn\\ 
& + \frac{m_4^4}{4(m_4^2-m_1^2) (m_4^2-m_2^2) (m_4^2-m_3^2)} \log \left(\frac{m_1^2}{m_4^2}\right)\,. 
\end{align}

The loop functions entering the approximate expressions are given by:
\begin{align}
D_0(z,t) &=  \frac{t \log t}{(1-t)^2 (z-t)} - \frac{1}{(t-1)(z-1)} 
+\frac{z\log(z)}{(t-z)(1-z)^2}\,, \\
D_2(z,t) &= \frac{t^2 \log t}{4(1-t)^2 (z-t)} - \frac{1}{4(t-1)(z-1)} 
+\frac{z^2\log(z)}{4(t-z)(1-z)^2}\,,
\\
D_0(x) &=  \frac{-2 + 2x - (1 + x) \log(x)}{(1-x)^3} \,, \\ 
D_2(x) &=  \frac{-1 + x^2 - 2x \log(x)}{4(1-x)^3}\,,
%D_0(1) &=& 1/6, \, \, \, \, D_2(1) = -1/12
\end{align}
\begin{align}
f_6(x,y) &=  \frac{x^2y^2-5x^2y-5xy^2-2x^2-2y^2+10xy+7x+7y-11}{6(x-1)^3(y-1)^3} \nonumber\\
& - \, \frac{y\log(y)}{(y-x)(y-1)^4} + \, \frac{x\log(x)}{(y-x)(x-1)^4}\,,
\end{align}
\begin{align}
\tilde{f}_6(x,y)  &=
-\frac{2x^2y^2+5x^2y+5xy^2-x^2-y^2-22xy+5x+5y+2}{6(x-1)^3(y-1)^3}    \nonumber\\
&- \, \frac{y^2\log(y)}{(y-x)(y-1)^4}        + \, \frac{x^2\log(x)}{(y-x)(x-1)^4} \,,
\end{align}

\begin{eqnarray}
       f_6(x)  &=& \frac{6 (1 + 3x) \log(x) + x^3 - 9 x^2 - 9x + 17 }{6(x-1)^5}\,, \\
\tilde{f}_6(x) &=& \frac{6x (1 + x)\log(x) - x^3 - 9x^2 + 9x + 1}{3(x-1)^5}\,.  
%\\
%f_6(1) &=& 1/20, \, \, \, \, \tilde{f}_6(1) = -1/30
\end{eqnarray}
%

%%%%%%%%%%%%%%%%%%%%%%%%%%%%%%%%%%%%%%%%%%

%%%%%%%%%%%%%%%%%%%%%%%%%%%%%%%%%%%%%%%%%%%%%%%%%%%%%%%%%%%%%%%%%%%%%%%%%%%%%%%%%%%%%%%%%%%%%%%%%%%
\end{document}